\documentclass{article}

\usepackage{Support_Docs/PRIMEarxiv}

\usepackage{amsmath,amsfonts,bm}

\def\eqref#1{equation~\ref{#1}}

\def\1{\bm{1}}

\def\va{{\bm{a}}}
\def\vb{{\bm{b}}}

\def\vg{{\bm{g}}}

\def\vx{{\bm{x}}}

\def\mI{{\bm{I}}}

\def\mW{{\bm{W}}}

\DeclareMathAlphabet{\mathsfit}{\encodingdefault}{\sfdefault}{m}{sl}
\SetMathAlphabet{\mathsfit}{bold}{\encodingdefault}{\sfdefault}{bx}{n}

\usepackage[utf8]{inputenc} 
\usepackage[T1]{fontenc}    
\usepackage{hyperref}       
\usepackage{url}            
\usepackage{booktabs}       
\usepackage{amsfonts}       
\usepackage{nicefrac}       
\usepackage{microtype}      
\usepackage{lipsum}
\usepackage{fancyhdr}       
\usepackage{graphicx}       
\usepackage{cleveref} 
\usepackage[backend=biber,natbib=true,style=chicago-authordate,]{biblatex}
\usepackage{dsfont}
\usepackage{comment}
\usepackage[dvipsnames,table,xcdraw]{xcolor}
\usepackage{amsmath}
\usepackage{enumitem}
\usepackage[most]{tcolorbox}
\usepackage{wrapfig}

\makeatletter
\AddToHook{cmd/appendix/before}{\def\cref@section@alias{appendix}}
\makeatother

\newcommand{\will}[1]{\textcolor{blue}{[\textit{Will: #1}]}}

\newcounter{Phenomena}[section]

\def\namedlabel#1#2{\begingroup
    #2%
    \def\@currentlabel{#2}%
    \phantomsection\label{#1}\endgroup
}

\newtcolorbox[auto counter]{boxy}[1][]{
  enhanced,
  breakable,
  fonttitle=\scshape,
  #1,
  before upper={\parindent15pt},
  boxrule=0pt,
  colback=white,
  borderline west={2pt}{-1pt}{white},
  borderline east={2pt}{-1pt}{white},
  borderline south={2pt}{-1pt}{white},
}
  
\title{if grid cells are the answer, what is the question?\\ a review of normative grid cell theory}

\author{
  William Dorrell \\
  Gatsby Unit, London \\
  \texttt{dorrellwec@gmail.com}
  \And James Whittington \\
  Oxford University
}

\begin{document}
\maketitle

\begin{abstract}
For 20 years the beautiful structure in the grid cell code has presented an attractive puzzle: what computation do these representations subserve, and why does it manifest so curiously in neurons. The first question quickly attracted an answer: grid cells subserve path-integration, the ability to keep track of one's position as you move about the world. Subsequent work has only solidified this link: bottom-up mechanistic models that perform path-integration match the measured neural responses, while experimental perturbations that selectively disrupt grid cell activity impair performance on path-integration dependent tasks. A more controversial area of work has been top-down normative modelling: why has the brain chosen to compute like this? Floods of ink have been spilt attempting to build a precise link between the population's objective and the measured implementation. The holy grail is a normative link with broad predictive power which generalises to other neural systems. We review this literature and argue that, despite some controversies, the literature largely agrees that grid cells can be explained as a (1) biologically plausible (2) high fidelity, non-linearly decodable code for position that (3) subserves path-integration. 
As a rare area of neuroscience with mature theoretical and experimental work, this story holds lessons for normative theories of neural computations, and on the risks and rewards of integrating task-optimised neural networks into such theorising.
\end{abstract}

\section{Introduction}

\begin{figure}[h]
    \centering
    \includegraphics[width=0.8\linewidth]{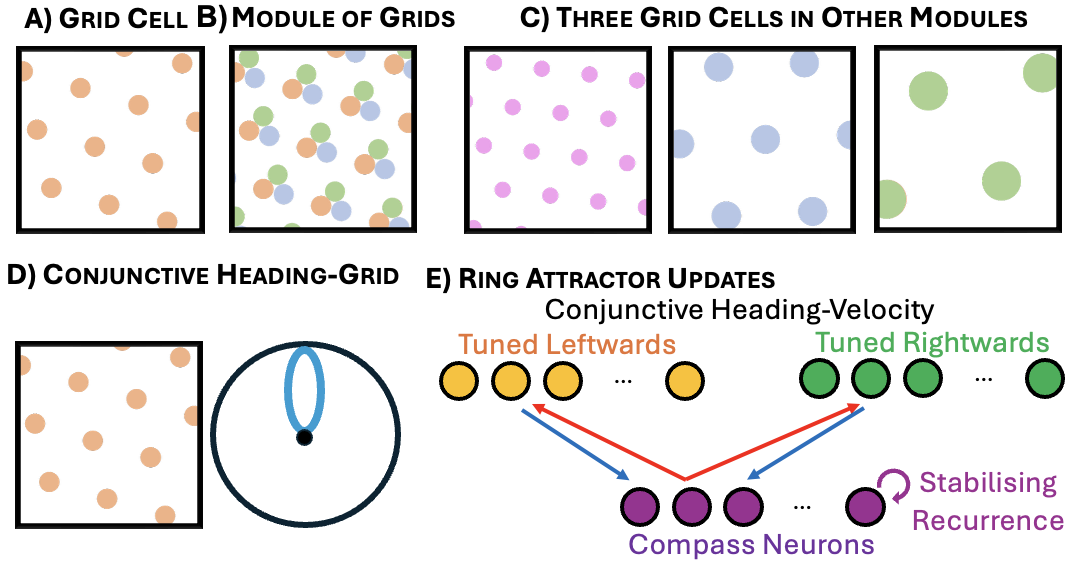}
    \caption{\textbf{Structure of the grid cell code. A:} Neurons are tuned to a hexagonal lattice of positions in 2D space. \textbf{B:} They are grouped into modules: neurons in the same module have translated (but not rotated) receptive fields, and across a module they uniformly sample the phases (translations). \textbf{C:} There are only a handful of modules in one animal, each with its own lattice, and $\sim$ 1000s neurons covering the possible phases. \textbf{D:} For each grid module there is a population of grid cells that are conjunctively tuned to both the underlying grid of the module, and a particular heading direction. \textbf{E:} These conjunctive neurons can implement path-integration by pushing the bump of neural activity around the module \citep{burak2009accurate}, like the ring attractor in the fly central complex \citep{hulse2020mechanisms}, using a shifted connectivity pattern: pure spatial neurons project to conjunctive neurons with the same spatial tuning profile (red connections), which project back to the spatial neurons shifted by their velocity tuning (blue connections). When the rightward neurons are more active than the leftward, this will cause the activity bump to move rightwards on the ring, implementing path-integration.}
    \label{fig:grid_lit-schematic}
\end{figure}

It has been 20 years since the discovery of the most surprising single neuron response yet described: grid cell activity correlates with an animal's self-position, activating when the animal is in a hexagonal lattice of positions \citep{hafting2005microstructure},~\cref{fig:grid_lit-schematic}A. Perhaps even more surprising than their original discovery is the finding that the grid cells lattices come in discrete modules of which a rodent will have a handful~\citep{stensola2012entorhinal},~\cref{fig:grid_lit-schematic}C. Grid cells in the same module have receptive fields that are translated (but not rotated) versions of one another which uniformly tile the space of possible phases,~\cref{fig:grid_lit-schematic}B. Finally, alongside the grid cells in layer II of medial entorhinal cortex, layer III hosts cells that fire at conjunctions of a hexagonal lattice of positions and a particular heading direction~\citep{sargolini2006conjunctive},~\cref{fig:grid_lit-schematic}D. There exists extensive additional phenomenology; but these four phenomena form a cohesive explanatory target:
\begin{enumerate}[label=P\arabic*., ref=P\arabic*]
    \item\label{itm:P1} Hexagonal-lattice tuning curves
    \item\label{itm:P2} For each grid cell there is a family of grid cells, called a module, which share the same tuning curve but translated, tiling the whole space.
    \item\label{itm:P3} The grid cell code contains multiple modules with different lattices.
    \item\label{itm:P4} The existence of paired conjunctive grid-heading direction cells.
\end{enumerate}
\begin{center}
\it{Giving these striking findings our questions are clear: what do grid cells do? And why in this way?}
\end{center}

A large body of work has convincingly answered the first question: the grid cell representation subserves path-integration. It has long been posited that the mammalian brain is capable of integrating its velocity to track self-position~\citep{tolman1948cognitive}, and as soon as grid cells were discovered they became the likely neural implementation~\citep{mcnaughton2006path}. In the intervening time the evidence has only built.

The second question is normative: why has biology chosen to perform path-integration using grid cells? Answering this question does not just satisfy curiosity; it promises principles to predict grid cell behaviour in novel situations, and the possibility that the same principles will generalise to other neural circuits. With the wealth of careful evidence that has accumulated the normative question seems well-posed and tractable. Despite this, there has been significant controversy in the field, producing a menagerie of different models whose commonalities and relative advantages are unclear.

This review seeks to clarify the normative grid cell theory literature. We proceed as follows:
\begin{enumerate}
    \item We begin with path-integration. We recall perturbative and mechanistic evidence that links grid cells to path-integration. Then we intuitively link the existence of translated tuning curves, \ref{itm:P2}, to path-integration.
    \item We then describe non-path-integrating `efficient coding' theories that model grid cells as only a high-quality positional encoding, not as position codes that connect to one-another via path-integration. We contrast with some natural instantiations of efficient coding for which place cells rather than grid cells are optimal. Then we show that those efficient coding approaches that do generate hexagonal tuning curves, \ref{itm:P1}, are unable to match the modular structure: sets of grid cells with translated axis-aligned tuning curves, \ref{itm:P2}. We justify this by explaining how this feature is detrimental for an efficient code, but crucial for path-integration.
    \item Next, we describe models that combine efficient coding with path-integration, and show that many classes of such models {\it are} able to capture the translated, axis-aligned, structure of grid cells, though most are limited to a single module. Further, we discuss the precise velocity update mechanism and the discrepancies between normative models and biology, in particular, \ref{itm:P4}.
    \item Finally, we discuss how nonlinear encoding objectives differ qualitatively from linear. Only with a nonlinear objective, along with a path-integration constraint, do multiple modules of grid cells appear, matching data, \ref{itm:P3}.
    \item We conclude with a unified normative view: theories that combine path-integration, nonlinear position encoding, and efficiency in the form of biological constraints (usually synaptic or neural activity energy efficiency and nonnegative firing rates) can cohesively capture \ref{itm:P1}, \ref{itm:P2} and \ref{itm:P3}: multiple axis-aligned grid modules. Further, we sketch remaining puzzles, including regarding \ref{itm:P4}, and lessons for the future.
\end{enumerate}

\section{Grid Cells Perform Path-Integration}

In this section we link the existence of a translated set of tuning curves, \ref{itm:P2}, to path-integration. We begin by reviewing evidence that grid cells are involved in path-integration. We then sketch intuitively how a translated set of tuning curves can naturally underlie path-integration.

\subsection{Non-Normative Evidence Linking Grid Cells to Path-Integration}\label{sec:non-normative_evidence}

In this section we briefly review two of the key strands of evidence that suggest grid cells subserve path-integration: mechanistic models and perturbation effects.

\paragraph{Mechanistic Models} Mechanistic models that perform path-integration match neural observations. The most successful of these are continuous attractor neural networks (CANNs). CANNs were originally developed to model path-integration of heading direction \citep{skaggs1994model,redish1996coupled}. Their simplest implementations comprise one population of neurons that encode the animal's heading direction, and two further populations that code for conjunctions of heading direction and angular velocity, either to the left or right,~\cref{fig:grid_lit-schematic}E. These conjunctive heading-velocity neurons can then be used to update the heading direction representation. First theoretically posited in the 90s, these circuits have since been verified experimentally, most beautifully in the fruit fly~\citep{kim2017ring}.

Subsequent work extended CANNs to two-dimensional space, initially to model hippocampal place cells~\citep{touretzky1996theory,samsonovich1997path,conklin2005controlled}. One difficulty in moving from a compact space of heading directions to an infinite space of (2D) positions is encoding the space in a finite set of neurons. Work that predated the discovery of grid cells proposed encoding space periodically, predicting lattice tuning curves but with square rather than hexagonal lattices~\citep{samsonovich1997path}. Subsequent work has shown how attractor dynamics in these 2D continuous attractor circuits can naturally leads to hexagonal grid and conjunctive cells~\citep{fuhs2006spin,guanella2007model,pastoll2013feedback,burak2009accurate}, and multiple modules~\citep{kang2019geometric,khona2025global}.

\ref{itm:P4}, the layer III conjunctive neurons, provide crucial evidence for these models. In a CANN each pure grid cell (i.e. tuned only to space) excites a set of conjunctive grid cells which have the same spatial tuning curve,~\cref{fig:grid_lit-schematic}E, but are additional tuned to movement in particular direction,~\cref{fig:grid_lit-schematic}D. In a CANN these cells implement path-integration by projecting back to the pure grid cell whose receptive field is translated along the direction of motion tuning,~\cref{fig:grid_lit-schematic}E. Not only do these modelled cells match those observed in layer III, but, remarkably, measured connections between layer II and III neurons estimated from spike-time connectivity match the shifted projection pattern \citep{vollan2025left}, presenting a ringing endorsement for the model.

There are other mechanistic models, notably the oscillatory-inteference model \citep{burgess2007oscillatory,burgess2008grid,bush2014hybrid,giocomo2008computation,hasselmo2008grid}. These models were motivated by the strong theta-frequency effects in entorhinal, including grid-cell phase precession \citep{hafting2008hippocampus,reifenstein2012grid}. However, they are unable to explain the presence of conjunctive grid cells, and more recent versions of CANN models that include theta-modulations can explain frequency effects like phase precession and theta sweeps \citep{vollan2025left}. As such, there is strong mechanistic evidence that circuits supporting path-integration can match the measured biological effects.

\paragraph{Perturbation Effects} Concurrently, behavioural evidence has shown that perturbing the grid cell system impairs animals' ability to perform path-integration dependent tasks. First, lesions to the medial entorhinal cortex impair path-integration~\citep{van2013distinct, steffenach2005spatial}. Second, disrupted spatial navigation is a known symptom of Alzheimer's disease, and this effect is thought to arise due to disruptions in grid coding in the medial entorhinal cortex. Evidence comes from genetic knock-in models of Alzheimer's which have disrupted grid cells~\citep{jun2020disrupted,ying2022disruption}, alongside impaired path-integration abilities~\citep{ying2022disruption}. Further, people at genetic risk of Alzheimer's show disrupted grid coding long before displaying other symptoms of Alzheimer's~\citep{kunz2015reduced}. Finally, and most precisely, removal of NMDA glutamate receptors from retro-hippocampal regions led to a selective disruption of grid cells while leaving other spatially selective cells intact. This perturbation caused behavioural disruptions to path integration~\citep{gil2018impaired}. In sum, the behavioural evidence is specific and strong.

\subsection{An Intuitive Guide to the Grid Cell Solution to Path-Integration}\label{sec:intuition}

We now outline how translational symmetry amongst tuning curves, \ref{itm:P2}, forms a natural substrate for path-integration. For simplicity, we work here with binary neurons that are either on or off, but the arguments generalise.

\begin{figure}
    \centering
    \includegraphics[width=0.65\linewidth]{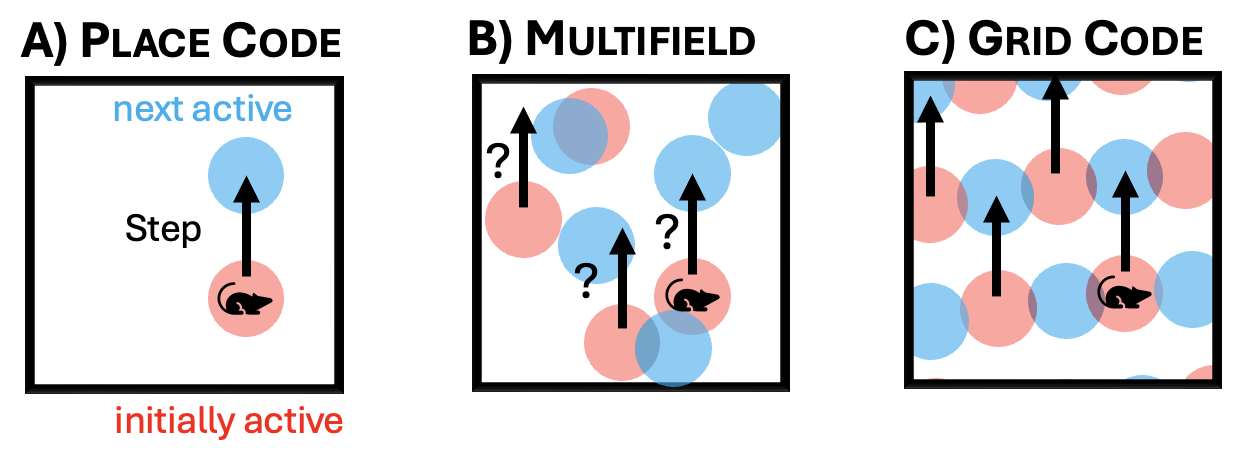}
    \caption{\textbf{Path-integration with different codes} \textbf{A:} Path-integrating with a place cell code is easy, current cell plus step uniquely determines next cell, but it is limited by the number of cells. \textbf{B:} Multifield cells improve the coding capacity but make path-integration more challenging, instead resources must be devoted to learning a mapping between unique combinations of cells. \textbf{C:} Within a grid moudle, current cell plus movement again uniquely determines the next cell: no matter which firing field of a grid cell you are in, thanks to the translational symmetry, you always know which cell to activate after a step. As such, grid cells elegantly combine the easy path-integration of place cells, with the higher capacity coding of multifield cells, and the path-integration mechanism generalises across space.}
    \label{fig:path_int_schematic}
\end{figure}

Path-integration involves updating your representation in response to movement. Upon taking a step, $\Delta \vx$, you have to update your internal encoding of position, $\vg(\vx)$, appropriately:
\begin{equation}
    \vg(\vx) \rightarrow \vg(\vx + \Delta \vx)
\end{equation}
A place cell code would make such updates very easy. The combination of the currently-active place cell and your movement specify the next representation: the place cell displaced by the movement,~\cref{fig:path_int_schematic}A. However, this requires a place cell for every potential position, limiting how many positions you can encode.

Instead imagine a cell that activates in multiple positions---a multifield place cell,~\cref{fig:path_int_schematic}B. These neurons can improve your encoding of position: rather than giving each position a unique cell, they are given a unique combination of cells, of which there are many more, improving the capacity of the code. However, this implies a more complex path-integration mechanism: knowing that a neuron is active and which movement you make is not enough; you need to know the full set of currently active neurons, and, upon stepping north, must have a mechanism to map each combination to its neighbour one step north. This, while possible, is much more complex and specific only to the particular arrangement of firing fields.

Modules of grid cells combine the coding quality of multifield cells with simple path-integrability \citep{kubie2012linear}. Each position is encoded by a combination of neurons, one in each module, leading to a more informative multifield-like code. Crucially, however, the path-integration problem is separated by modules, and within each module it is simple. Knowing that one neuron in a module is active and that you make a movement north uniquely determines which neuron in that module should be active next---the one with a receptive field translated one step north,~\cref{fig:path_int_schematic}C. By baking translational symmetry into the multifield pattern path-integration is made easy.

In short, these are the functional insights that underlie the grid cell code: a dense multifield code for position combined with easy module-wise path-integration. Indeed, in the final section, we conclude by outlining how a combination of these two functional goals with simple biological considerations (nonnegative small firing rates) leads to grid cells. For now we turn to attempts to model grid cells without reference to path-integration.

\section{Grid Cells are \textit{not} the most Efficient Code for Space}\label{sec:grid_lit_efficient_coding}

In the previous section we outlined the links between path-integration and grid cells, in particular their modular translated receptive-field structure, \ref{itm:P2}. In contrast, in this section we review what we term `efficient coding' theories of grid cells. These normative models posit that grid cells are the most efficient encoding of position, without mentioning path-integration. We labour on these models as many have become prevalent, yet they lack the key computational feature that defines entorhinal cortex---path-integration---and do not match many critical aspects of grid cell data. 
We will begin by showing instantiations of efficient coding that do not generate hexagonal tuning curves. We will then discuss efficient coding that do generate hexagonal tuning curves, \ref{itm:P1}, but will show that in each case they do not capture the translated receptive fields, \ref{itm:P2}, a symptom of dropping path-integration.

\subsection{Context: Many Efficient Coding Models do not generate Grid Cells}\label{sec:grids-lit_sengupta}

Most efficient coding theories can be decomposed into two parts. The first measures the quality of the encoding, for example, how well can a linear decoder predict where you are from your representation. The second measures or enforces the efficiency or biological plausibility of the code, for example via low nonnegative firing rates. Combinations of the two lead to some of the famous results in theoretical neuroscience, such as histogram equalisation via the fly eye's nonlinearity \citep{laughlin1981simple}, whitening via centre-surround in retinal ganglion cells \citep{atick1990towards}, or sparsifcation of natural images via the V1 gabor code \citep{olshausen1996emergence}. 

Before studying efficient coding theories that generate grid cells, we make a useful counterpoint: very natural instantiations of efficient coding of space do not produce grids. Comparing between these theories clarifies the choices that lead to grids. \citet{sengupta2018manifold} use the similarity matching objective: given two inputs (e.g. positions), $\vx$ and $\vx'$, and their neural encodings, $\vg(\vx)$ and $\vg(\vx')$, this objective encourages the dot-product similarity of the representation, $\vg(\vx)^T\vg(\vx')$, to match that of the input similarity structure, $\vx^T\vx$, through maximising the following loss:
\begin{equation}\label{eq:sim_match}
    \mathcal{L} = \iint \underbrace{(\vx^T\vx' - \alpha)}_{\text{Input Similarity}}\underbrace{\vg(\vx)^T\vg(\vx')}_{\text{Representation Similarity}} \underbrace{dp(\vx)dp(\vx')}_{\text{Occupancy probability}} \qquad \text{subject to} \qquad \underbrace{||\vg(\vx)|| = 1}_{\text{Normed activity}},\quad \underbrace{\vg(\vx)\geq 0}_{\text{Nonnegativity}}
\end{equation} 
\cite{sengupta2018manifold} take inputs from a compact continuous space, such as angles on a ring, and (reasonably) assume that the input similarity, $\vx^T\vx$, decays with distance: nearby points are similar, distant are dissimilar. From this they analytically derive that, with infinitely many neurons, place cells are the optimal nonnegative representation. This is not specific to this loss: recent work has drawn similar conclusions from an information theoretic measure of coding quality \citep{deighton2024higher}. This is somewhat natural, place cells are a very informative code, and a much simpler one than multifield codes. When there are enough neurons such that a place cell code can tile the space with sufficient resolution, these works present evidence that some efficient coding approaches prefer place cells (in~\cref{sec:nonlinear_decoding} we also show that place cells are preferred even with few neurons).

As such, it seems difficult for efficient coding of space alone to produce grid cells. To modify an efficient coding theory we can either change how coding quality is measured or the efficiency constraints. Many efficient coding models can be described in this way and succeed in generating hexagonal lattice tuning curves, \ref{itm:P1}. They are, however, unable to account for each module's axis-aligned translated receptive field structure, \ref{itm:P2}. We conceptually cluster these approaches into two groups, nonnegative bandpass filter models, which we review next, and clustering models, which we review in~\cref{app:dense_packing}.

\subsection{Grid Cells via Nonnegative Bandpass Filtering}\label{sec:bandpass_filters}

We now review nonnegative efficient coding grid cell models that generate hexagonal lattices via nonnegative Fourier combinations, and in particular, a bandpass filter effect. These include nonnegative PCA models \citep{dordek2016extracting,sorscher2019unified,sorscher2023unified} and metric encoding models \citep{pettersen2024self}.

\paragraph{Nonnegative PCA of difference-of-Gaussian Place Cells} The first set of models use an encoding objective that rewards the representation for containing high power at a critical spatial frequency, then use nonnegativity to produce a hexagonal lattice. The pivotal link in these arguments was first described by \citet{dordek2016extracting} who modelled grid cells as the nonnegative PCA of difference-of-Gaussian place cells, producing hexagonal receptive fields. This link is neat, but, in brief, it suffers from two major flaws. First, it relies on the use of difference-of-Gaussian place cells which are not observed; second, it fails to produce modules of translationally-symmetric grid cells.

\begin{figure}[h]
    \centering
    \includegraphics[width=0.9\linewidth]{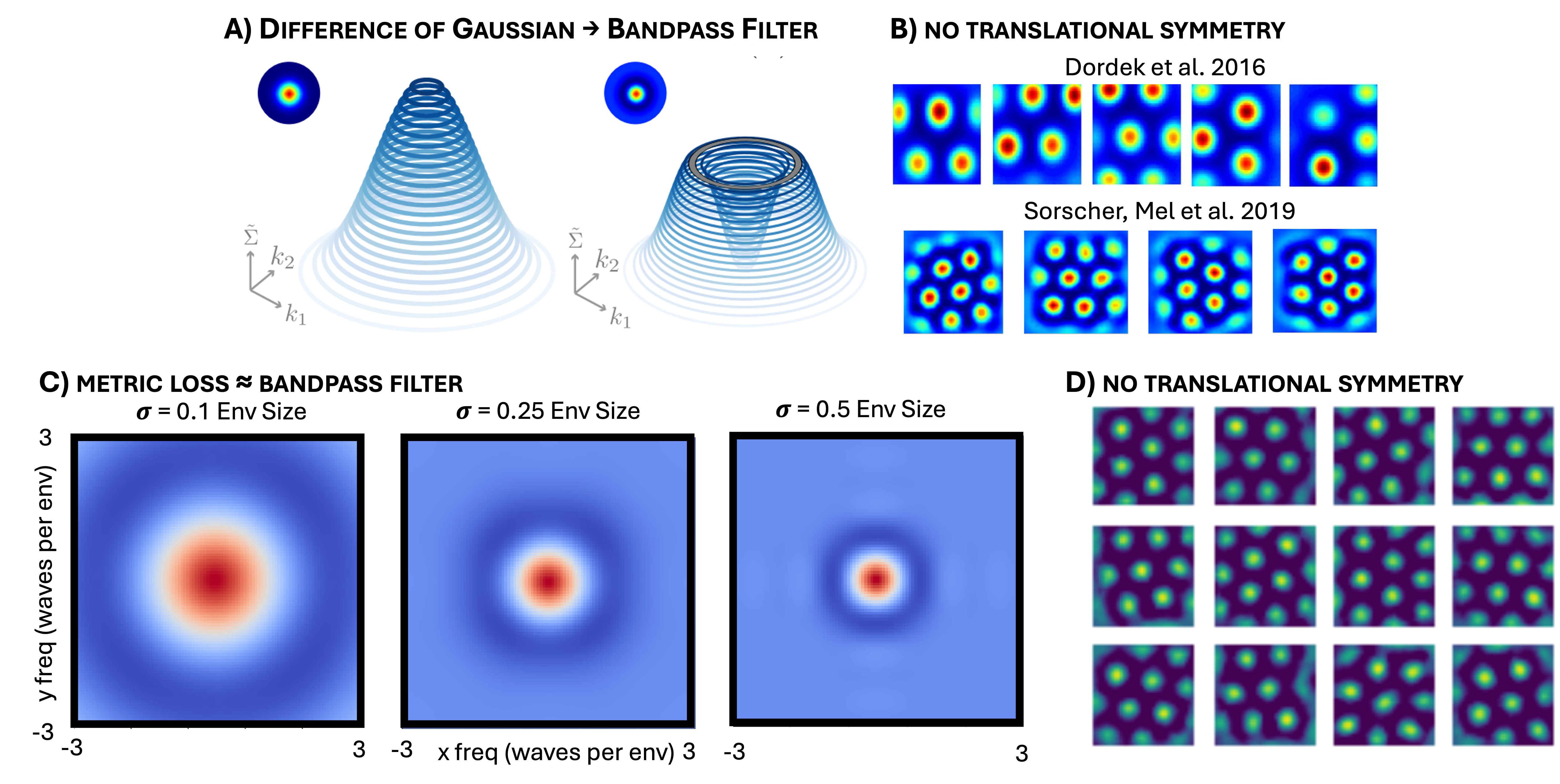}
    \caption{\textbf{Grid Cells via Bandpass Filtering. A:} A Gaussian place cell code has a covariance whose frequency content is a smoothly-decaying Gaussian, left, but a difference-of-Gaussian code has covariance whose frequency content peaks at a non-zero frequency, figure from \citet{sorscher2019unified}. \textbf{B:} The grid cells that result from nonnegative PCA on difference-of-Gaussian place cells are not translationally symmetric, each population contains grid cells whose axes are rotated relative to one another (for example, the left and rightmost grid cells from dordek have lattices rotated 30$^\circ$ relative to one another), figures from \citet{dordek2016extracting,sorscher2019unified}. 
    \textbf{C:} We create a representation, $\vg(\vx)$, that contains a single frequency, and plot the conformal loss,~\cref{eq:pettersen}, as a function of this single frequency for a few $\sigma$ values. This loss is minimised (dark blue) at an intermediate value of frequency: a bandpass filtering effect. \textbf{D:} Metric encoding also produces a population of grid cells that are rotated relative to one another, figure from~\citep{pettersen2024self}.}
    \label{fig:fourier_models}
\end{figure}

The similarities to the approaches in~\cref{sec:grids-lit_sengupta} are large; the largest difference is the choice of target, $\vx$. rather than something like Gaussian place cells, whose similarity structure decays with distance, they use difference-of-Gaussian cells. \citet{dordek2016extracting} (later paralleled by \citet{sorscher2019unified,sorscher2023unified}) nicely explain the effect of this substitution: difference-of-Gaussian cells lead to a bandpass covariance structure peaked at a particular frequency band~\cref{fig:fourier_models}A, leading the optimal linearly-decodable representation to highly encode this frequency. Combining this with a lattice discretisation effect from the finite room leads to square grid cells \citep{dordek2016extracting}. Finally, enforcing nonnegative firing rates changes the optimal solution from square to hexagonal grids, justified either through a triplet interaction effect \citep{sorscher2019unified,sorscher2023unified}, or the efficiency in positivising the code \citep{dordek2016extracting}.

This approach has been influential with many papers using the nonnegative PCA of difference-of-Gaussian place cells \citep{dordek2016extracting,sorscher2019unified,sorscher2023unified,schoyen2023coherently,tang2024learning}. It has also been controversial, prompting a rebuttal \citep{schaeffer2022free}, a rebuttal to the rebuttal \citep{sorscher2022and}, and two further rebuttals cubed \citep{schaeffer2023disentangling,schaeffer2023testing}. One point of disagreement lay in the finetuning of parameters required to produce grid cells: an interesting point, but clearly not fatal since the brain could simply use these parameters. A more existential threat comes from the choice of difference-of-Gaussian tuning curves. These fit hippocampal place cells less well than Gaussian curves, but, as the theoretical analysis states, are clearly vital for the production of hexagonal grid cells. Many more realistic choices of place cells don't produce grid cells in this framework \citep{schaeffer2023disentangling}, since they don't generate the required bandpass filter. This could be an interesting prediction about the relationship between place and grid coding, but currently there's no evidence this particular link exists.

Second, and fundamentally, these approaches do not capture the translated receptive field structure of grid modules. Instead, they produce grid cells whose orientations cluster into two groups offset at 30 degrees \citep{pettersen2024self}~\cref{fig:fourier_models}B, a pattern that is not observed experimentally. Further, when they do produce multiple modules, the intermodule relationship appears to be worryingly governed by numerical discretisation effects \citep{sorscher2019unified}, nor does the framework offer an explanation of conjunctive cells, P4. Only when combined with a path-integrating task (for example by training an RNN to both path-integrate and linearly project to difference-of-Gaussian place cells) do you get axis-aligned grid cells, a topic we'll return to. Hence, this theory appears to be, at best, part of the solution.

\paragraph{Metric Encoding} A seemingly-distinct class of theories study a loss that encourages the `neural metric' to match the metric of space. We will show that we can understand these as performing a similar bandpassing effect as discussed.

A metric is a function that measures distances between points. Matching a particular metric means that the distance between two points, $\vx$ and $\Delta\vx$, is preserved in the distance between the representation of those points, $\vg(\vx)$ and $\vg(\Delta\vx)$, at least for a small region of space (small $\Delta\vx$):
\[|||\vg(\vx + \Delta\vx) - \vg(\vx)|| = s||\Delta\vx|| + \mathcal{O}(\Delta\vx^2)\]
where $s$ is a scaling factor. Normative approaches including losses like these are common routes to grid cells often in combination with path-integration~\citep{gao2019learning,gao2021path,xu2025conformal,pettersen2024self}. Here we focus on the findings of \citet{pettersen2024self}: optimising a nonnegative unit-norm representation to preserve distances while penalising the L1 norm of the firing rates is sufficient to generate hexagonal firing fields without path-integration. The loss used is:
\begin{align}\label{eq:pettersen}
    \mathcal{L} &= \overbrace{\alpha\mathbb{E}_{\vx,\vx'}\bigg[ e^{-\frac{1}{2\sigma^2}||\vx-\vx'||_2^2}(||\vx-\vx'||_2 - ||\vg(\vx)-\vg(\vx')||_2)^2\bigg]}^{\text{Conformal Isometry $\approx$ Similarity Matching}} - \overbrace{(1-\alpha)\mathbb{E}[||\vg(\vx)||_1}^{\text{L1 Capacity Loss}}] \\ \qquad &\text{subject to} \qquad ||\vg(\vx)|| = 1, \quad \vg(\vx)\geq 0
\end{align}
The first term, called the conformal loss, forces the neural distance, $||\vg(\vx)-\vg(\vx')||_2$, to match the separation in space, but only when $\vx$ and $\vx'$ are close, via the $e^{-\frac{1}{2\sigma^2}||\vx-\vx'||_2^2}$ weighting term. As such, it is conceptually close to similarity matching,~\cref{sec:grids-lit_sengupta}. In particular, the weighting sets a lengthscale, $\sigma$, on the local region in which similarity matching has to occur. If $\sigma$ is much larger than the environment, $ e^{-\frac{1}{2\sigma^2}||\vx-\vx'||_2^2} \approx 1$, the loss becomes a similarity matching one, and place cells are again the optimal representation with many neurons, as in \citet{sengupta2018manifold},~\cref{fig:Pettersen_Largesig_Place}.

When $\sigma$ is smaller this loss generates hexagonal grids. We now show that this can also understood as a Fourier bandpass effect. The loss contains two biases, one that penalises high frequencies, another low frequencies, that together create a bandpass filter. The local region, encapsulated by $\sigma$, sets a lower bound on the frequency content of the code: if your code contains a component oscillating slower than $\sim \frac{1}{\sigma}$ it won't have varied meaningfully within the regions you care about, so won't decrease the loss. Conversely the similarity matching part, $(||\vx-\vx'||_2 - ||\vg(\vx)-\vg(\vx')||_2)^2$, sets a high-frequency cutoff: the code should contain low frequencies so that nearby points are similar, and distant ones are different. We illustrate this for a neural code containing a single frequency by plotting the loss as a function of this frequency~\cref{fig:fourier_models}D. The loss is minimised at a particular frequency ring (shown in dark blue) whose radius scales with the inverse of $\sigma$. This is exactly the same bandpass filter of \citep{sorscher2019unified}.

Having established the bandpass filter, similar arguments to the previous section can then be used to justify how positivity and capacity constraints might lead to grid cells. Indeed, hexagonal grid cells with a single lengthscale emerge from this optimisation, with the lengthscale controlled by $\sigma$ \citep{pettersen2024self}. This is not a complete picture: for example, it is an interesting mathematical puzzle that combining this loss with an L1 capacity constraint, but not an L2, leads to hexagonal grids \citep{pettersen2024self}. Regardless, these grid cells still suffer from the same shortcoming of other efficient coding only approaches: the grids are not aligned within the same module, rather, they feature the same loose 30$^\circ$ alignment as the Fourier approaches,~\cref{fig:fourier_models}E. Only by adding path-integration is this effect removed.

\paragraph{Summary} Nonnegative combinations of Fourier components can generate hexagonal grid cells. In addition to some plausibility concerns (place cells are not well modelled by difference-of-Gaussians), without path-integration, these models are unable to reproduce the translationally symmetric modular structure that is vital for path-integration.

\subsection{Conclusion: Inefficiency of Axis-Aligned Grid Cells} 

From this large body of work (see also clustering models in~\cref{app:dense_packing}) we conclude that grid cells, despite clearly being a good code, are not the optimal efficient code of 2D space. In natural instantiations of the efficient coding problem the optimal solution are place cells (with either one or multiple fields depending on the problem,~\cref{sec:nonlinear_decoding}). This matches unpublished findings from Tzushuan Ma's PhD thesis \citep{ma2020towards}, and recent work that shows multifield place cells, as in the hippocampus, are a very good code \citep{rich2014large,harland2021dorsal,eliav2021multiscale}. Changing the problem in various ways can make hexagonal-lattice receptive fields optimal, either through a bandpass filter,~\cref{sec:bandpass_filters}, or a dense packing argument,~\cref{app:dense_packing}. However, it never recovers translational symmetry. This is intuitive: the grid-cell code has some glaring design flaws from a pure efficient coding perspective. The periodicity of grid cells means they identically encode points separated by the lattice symmetry, rendering a single cell unable to distinguish them. The translational symmetry within a module means that rather than helping each other to decode new points, points that are indistinguishable to one neuron are also indistinguishable to all neurons in the module! Breaking the symmetry, either by rotating and scaling the grid lattices of different neurons or removing the lattice entirely, usually improves the coding quality. As such, translated receptive fields, \ref{itm:P2}, are a key symptom of grid cells' role in path-integration, and very hard to justify from an efficient coding perspective.

\section{Path-integration + Position Encoding = A Module of Grid/Place cells}

In~\cref{sec:intuition}, we outlined how grid modules' translational symmetry forms an ideal substrate for path-integration, something that purely efficient coding approaches are unable to capture. Here, we review various models that combine path-integration with an encoding loss and recover a single module of axis aligned grid cells.

\subsection{Path-Integrating Models of Grid Cells}\label{sec:path_integrating_theories}

\paragraph{Path-Integrable Efficient Codes} \citet{dorrell2023actionable}, similarly to unpublished work \citep{ma2020towards}, use mathematical analysis to combine path integration with the earlier efficient coding approaches. Identically to an efficient coding approach, the representation is asked to encode space subject to some efficiency constraints. However, crucially, the code is also asked to permit path-integration: $\vg(\vx + \Delta \vx) = f(\vg(\vx), \Delta \vx)$ predicting next representation, $\vg(\vx + \Delta \vx)$, from the current representation, $\vg(\vx)$, and velocity, $\Delta \vx$. For mathematical analysis, this constraint is enforced using action-dependent weight matrices: each weight matrix has to correctly implement all transformations of the code for a given action, independent of the animal's current position:
\begin{equation}\label{eq:actionability}
    \vg(\vx + \Delta \vx) = \mW(\Delta \vx) \vg(\vx) \quad \forall \vx
\end{equation}
This constraint ensures that if the agent is at a position $\vx$, it can use $\mW(\Delta \vx)$ to predict where it will reach next, permitting path-integration. Further, it can be mathematically derived that this constraint forces the code to contain a small number of Fourier features, providing a basis for further analysis. Combining this with an efficient coding loss leads to either one or multiple modules depending on the choice of loss \citep{dorrell2023actionable}. It does not directly explain the conjunctive grid coding, nor are action dependent weight matrices particularly biologically plausible. Both of these problems can be alleviated through action gating, a plausible scheme to implement action-dependent weight matrices as seen in other models \citep{logiaco2021thalamic}. 

\paragraph{Efficient Coding of Trajectories} \citet{rebecca2025spatial}, following similar work by \citet{waniek2020transition}, formulate grid cells in a reversed manner: rather than requiring velocity to  update the encoding from one timestep to the next, they instead predict velocity from each current and next encoding. From this approach, and a small number of assumptions, they show that a single hexagonal grid module is optimal for predicting velocity. While elegant, this argument suffers from using binary neurons and a discretisation of space, and struggles to naturally encapsulate multiple modules. Regardless, this alternate formulation of path-integration makes some useful novel predictions, such as how a 2D module should encode a 1D sequence.
\begin{wrapfigure}[20]{r}{0.4\textwidth}
  \begin{center}
    \includegraphics[width=0.42\textwidth]{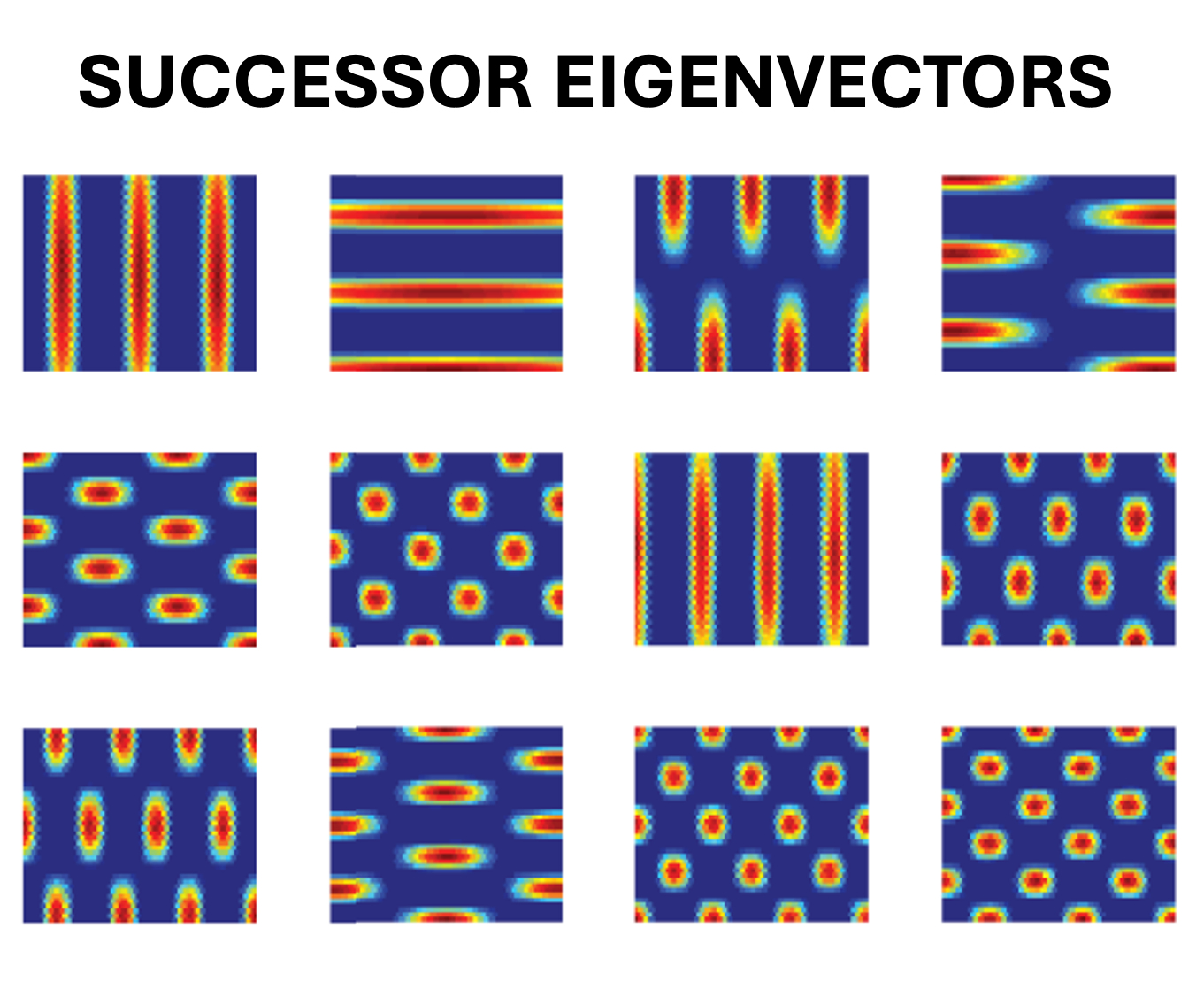}
  \end{center}
  \caption{Successor representation eigenvectors are poor models of grid cells, figure from~\citet{stachenfeld2017hippocampus}.}
  \label{fig:sucessor}
\end{wrapfigure}

\paragraph{Grids as Eigendecomposition of Transition Matrices} 
A set of models have formalised spatial coding via transitions on 2D graphs. For example, \Citet{stachenfeld2017hippocampus} argue that the hippocampus encodes a successor representation (a simple function of a transition matrix) of space, and that the thresholded-nonnegative eigenvectors of the successor representation (and thus the transition matrix)---which are periodic---correspond to grid cells. Later \citet{yu2020prediction} generalised this approach, showing that directed, rather than diffusive, transitions matrices can be used to path-integrate. However, the grid cells that emerge from eigendecomposition of such transition matrices are unlike real grid cells. They exist in modules of only two neurons, many of which are not hexagonal grids but instead form bands or amorphous blobs,~\cref{fig:fourier_models}C, especially in non-square rooms \citep{stachenfeld2017hippocampus}. Further, while one of the selling points of the successor representation theory is its sensitivity to transition statistics, pure grid cells only emerge with a diffusive policy, whereas real grid cells are more robustly hexagonal \citep{stensola2012entorhinal,vollan2025left}. Thus, while these models are an elegant mathematical framing, they leave several unanswered questions: why only some eigenvectors match grid behaviour; why each modelled grid module has only 2 neurons per module; why empirical grid cells are not so dramatically affected by transition statistics; and how this model could account for conjunctive grid cells. 

\paragraph{Neural Network Models} The most common path-integration approach is to train recurrent neural networks (RNN) to path-integrate, and then to use the learnt internal representation as a model of grid cells. In its simplest instantiation, RNNs are provided a sequence of actions, and required to output the corresponding sequence of positions. This captures all three aspects of the efficient path-integrating code above: the code must path-integrate, it must distinguish different points so they can be decoded, and it must do efficiently; with low weights (if using regularisations) and with nonnegative activities (if using ReLU nonlinearities). However, the precise design choices, and the results, have varied considerably.
\begin{itemize}
    \item Some models provide the action as a standard input to the RNN, $\va(t)$:
\begin{equation}\label{eq:standard_input_RNN}
    \vg(t+1) = \sigma(\mW\vg(t) + \mI\va(t) + \vb)
\end{equation}
while others learn a mapping between the action and the recurrent weight matrix, similar to the normative models above:
\begin{equation}\label{eq:action_depedent_RNN}
    \vg(t+1) = \sigma(\mW(\va(t))\vg(t) + \vb)
\end{equation}
\item Some networks predict $(x,y)$ coordinates, others Gaussian place cells or difference-of-Guassian place cells.
\item Some networks use a ReLU nonlinearity, enforcing nonnegativity, others use tanh. 
\item Weight or activity is often constrained, either through a regularisor, or through a unit norm constraint.
\item Other regularisors might be added, most often the conformal isometry loss,~\cref{sec:bandpass_filters}.
\end{itemize}

An early pair of results suggested that path-integrating RNNs could model grid cells. \Citet{cueva2018emergence} trained standard RNNs to path-integrate and found grid and band-like neurons, though these grids were often square rather than hexagonal. Key choices included the use of tanh rather than ReLU nonlinearity, meaning the activities were both positive and negative, and reading out $(x,y)$ coordinates rather than a place cell code. Concurrently, \citet{banino2018vector} trained a large reinforcement learning model and showed that a feedforward layer in the network, heavily regularised by dropout, learnt somewhat griddy neurons, though there are concerns that these `grid cells' are as grid-cell-like as low-pass filtered noise \citep{sorscher2019unified}.

Since then, the class of models that learn an action-dependent weight matrix,~\cref{eq:action_depedent_RNN}, have been very successful. First studied by \cite{issa2012universal}, who derived conditions for such a model to work, these were then used as part of a larger model of the hippocampal-entorhinal system by  \citet{whittington2018generalisation,whittington2020tolman}, who trained sub-networks to path-integrate, and found hexagonal modules of grid cells, though they baked the modular structure into the network. Another vein of work used the conformal isometry losses and a difference-of-Gaussian place cell readout to learn a single module of hexagonal grid cells~\citep{gao2019learning,gao2021path,xu2020theory}. Finally, \citet{schaeffer2023self} showed that training the action-dependent matrices in a ReLU RNN with a unit-norm constraint, an activity loss to reduce network capacity, a conformal loss, and a separation loss, led to multiple modules of axis aligned grid cells. Since these models do not explicitly capture the way velocity is coded by neurons, instead embedding it in the changing weight matrix, this architecture will never capture the conjunctive grid cells. Despite this, they present a ringing endorsement for the idea that optimising for a good, efficient, path-integrating code for position is sufficient for recovering grid-cells.

Path-integrating in more standard RNNs,~\cref{eq:standard_input_RNN}, can also lead to grid cells. \citet{sorscher2019unified,sorscher2023unified} trained such an RNN to predict difference-of-Gaussian place cells and found a single axis-aligned module of grid cells, later supported by \citet{tang2024learning}. A similar story was seen in \citet{pettersen2024self}, who showed that a metric approach combined with path-integration led to a single module of axis-aligned hexagonal grid cells. Finally, \citet{xu2025conformal} show that a standard RNN formulation with a unit-norm, positivity, and conformal constraint is sufficient to generate a single module of grid cells, matching theoretical work~\citep{schoyen2025hexagons}. Each of these approaches highlight a move from efficient coding-only approaches to path-integration: the coding losses alone produce hexagonal grid cells, but the axes of these grid cells are not aligned,~\cref{sec:bandpass_filters}. Additionally asking for path-integration aligns the axes.

Each of these models demonstrates that RNNs trained to path-integrate naturally generate a module of grid cells. We will focus on two further points of discrepancies. In~\cref{sec:nonlinear_decoding}, we will discuss how many of these models are limited to a single module. First, however, no model has reported the path-integration mechanism using conjunctive grid cells, \ref{itm:P4}, as in purely mechanistic models \citep{burak2009accurate}, a discrepancy we will discuss next. 

\subsection{A Velocity Update Puzzle}\label{sec:velocity_update}

In this section we review an ongoing puzzle regarding the precise grid cell velocity-update mechanism. In~\cref{sec:non-normative_evidence} we discussed the how the pre-eminent mechanistic models, CANNs, use conjunctive neurons to path-integrate, matching connectivity measurements \citep{vollan2025left}. Here, we outline a discrepancy between this and normative models.

Of the path-integrating theories listed in~\cref{sec:path_integrating_theories}, most do not comment on velocity-update mechanism. They either abstract away from this part of the model, or use an action-dependent weight matrix that muddies how such dependence arises. The only models which do include such effects are RNNs with standard updates,~\cref{eq:standard_input_RNN}.  
Surprisingly, \citet{schoyen2023coherently,pettersen2024self} found that such networks learn a population of band-like cells, and that these are the neurons that seem to do the work of performing path-integration---the network can path-integrate without the grid cells! This is in contrast to a CANN model in which the grid cells are vital for the path-integration. \citet{chu2025unfolding} elegantly explain this finding: in task-optimised RNNs the two-dimensional path-integration problem is effectively broken down into two one-dimensional problems. Along two directions a population of cells integrates motion using a standard ring attractor architecture and, due to their focus on one dimension, these cell's tuning curves resemble band cells. Then, since they are using a bandpass filter loss which specifically encourages the formation of grid cells~\cref{sec:bandpass_filters}, a module of axis-aligned grid cells is generated from the band cells.

As such, it seems that the brain and task-optimised RNNs with standard architectural choices use fundamentally different path-integration mechanisms. Resolving this discrepancy remains an open question.

\subsection{Conclusion: Path-Integration and Axis-Aligned Grid Cells}

Overall, it seems well established that RNNs optimised to perform a task that includes (1) path-integration, (2) encoding of position, and (3) biological constraints (mainly nonnegativity and low firing rates) robustly learn grid cells. However, as yet the precise structure of the set of necessary constraints is unclear, especially when using a more standard RNN architecture, and the discrepancy between velocity-update mechanisms remains puzzling.

\section{Only with Nonlinear Encoding are Multimodular/Combinatorial Solutions Optimal}\label{sec:nonlinear_decoding}

By encoding each position with a unique combination of cells, combinatorial codes achieve higher capacity than unimodal codes,~\cref{sec:intuition}. However, this comes at a trade-off in ease of decoding position from such a code. In particular, here we outline how `linear' approaches cannot make use of multi-field codes and instead prefer either place cells or one module of grid cells; only with more powerful `nonlinear' approaches do combinatorial multifield place or multimodular grid representations become optimal. Lastly, we provide a cohesive summary of the conditions in which grid cells are optimal positional representations---nonlinear efficient codes of path-integration---and review successes at predicting the optimal size and alignment of grid modules.

\subsection{Combinatorial Codes Require Nonlinearity}\label{sec:intuition_nonlinear}

Consider a population of $N$ binary neurons; assigning each position its own disjoint set of cells can encode at most $N$ positions, one per neuron. Alternatively, a combinatorial scheme which assigns each position a unique but overlapping set of cells can produce up to $2^N$ unique codes, enormously expanding the set of encodable positions. It is this basic fact that makes combinatorial positional codes, be that the apparently random multi-scale code in the hippocampus \citep{eliav2021multiscale} or the multimodular structure of grid cells, more effective.

Yet, using such a combinatorial code requires nonlinear processing. Imagine trying to decode whether or not you are in position $\vx$. In a simple place cell code this can be done linearly: simply check whether the place cell uniquely corresponding to $\vx$ is on or off. It's similarly easy to decode position in a rotation of a place cell code. But in a combinatorial code, $\vx$ corresponds to many place cells, and each place cell corresponds to many $\vx$. Decoding $\vx$ from a combinatorial code thus requires responding to a specific conjunction of place cells, and this is not something that a linear decoder can do. It requires nonlinearity.

\paragraph{`Functionally linear' losses prefer single grid modules.} Losses that rely on linear decoding of place cells, PCA of place cells, or linear similarity objectives, such as~\cref{eq:sim_match}, struggle to profit from multimodularity. Indeed in our previous work we demonstrated that losses that are a linear function of similarity, such as~\cref{eq:sim_match}, exhibit a failure mode: they encourage further distinguishing already well distinguished positions rather than those that are poorly distinguished. This representational pressure leads to place cells or single modules of grid cells, rather than a combinatorial code \citep{dorrell2023actionable}. This finding reflects a broader pattern: all prior works that use metric encoding or nonnegative PCA of difference-of-Gaussian place cells is similarly `functionally linear', and to the best of our knowledge, all works that combine such losses with path-integration lead to a single module \citep{sorscher2019unified,sorscher2023unified,tang2024learning,schoyen2023coherently,pettersen2024self}. We note that while some models do report multiple modules using these losses, they only do so by baking a multiple modular structure into the code to begin with \citep{gao2019learning,gao2021path,xu2025conformal}, i.e. multiple modules do not emerge as the optimal code.

\paragraph{`Functional nonlinearity' profits from multiple modules.} This failure model of linear losses motivated us to introduce the following `nonlinear' similarity matching objective~\citep{dorrell2023actionable}:
\begin{equation}\label{eq:non_linear_similarity}
    \mathcal{L} = \iint \underbrace{\chi(\vx, \vx')}_{\text{Input Similarity}}\underbrace{e^{-\frac{||\vg(\vx) -\vg(\vx')||_2^2}{2\sigma^2}}}_{\text{Nonlinear Representational Similarity}} \underbrace{dp(\vx)dp(\vx')}_{\text{Occupancy probability}} 
\end{equation} 
In this loss, if the representations of two points are already well distinguished ($\vg(\vx)$ and $\vg(\vx')$ are already further apart than $\sigma$), no further gain is achieved by distinguishing them further. Instead, the code focuses its efforts on distinguishing poorly distinguished points. This encourages the formation of combinatorial codes, which make best use of the available neurons. Indeed, we know of only two normative models that derive multiple translationally symmetric modules as the optimal solution, ours \citep{dorrell2023actionable} and \citet{schaeffer2023self}. Both use the nonlinear similarity matching objective we proposed,~\cref{eq:non_linear_similarity}.

In sum, we suggest that this division between `functionally nonlinear or linear' losses---which correspond to linear or nonlinear decodability of position---can neatly explain which approaches generate single or multiple modules, depending on whether the loss is flexible enough to take full advantage from a combinatorial code.

\subsection{The Interplay of Path-Integration, Nonlinear Decoders, and Resource Constraints}

We are now in a position to summarise the optimality of different spatial representations as a function of a small number of key modelling choices: linear versus nonlinear loss functions, whether path integration is required, and neural resource constraints (i.e., the number of neurons)\footnote{Throughout, we assume nonnegative neural activity with unit norm.}.

One initially surprising finding is that, when many neurons are available, place cells are optimal independent of other considerations. In~\cref{sec:grids-lit_sengupta} we related how place cells are the optimal nonnegative similarity matching code when there are more neurons than positions to be distinguished. We find that the same is true with a nonlinear similarity matching loss, and/or with an additional path-integration constraint (for example, by enforcing actionability,~\cref{eq:actionability} \citep{dorrell2023actionable}). We suggest this is because when there are enough neurons, simple place cell codes can tile the space at sufficient resolution.

\begin{figure}[th]
    \centering
    \includegraphics[width=0.5\linewidth]{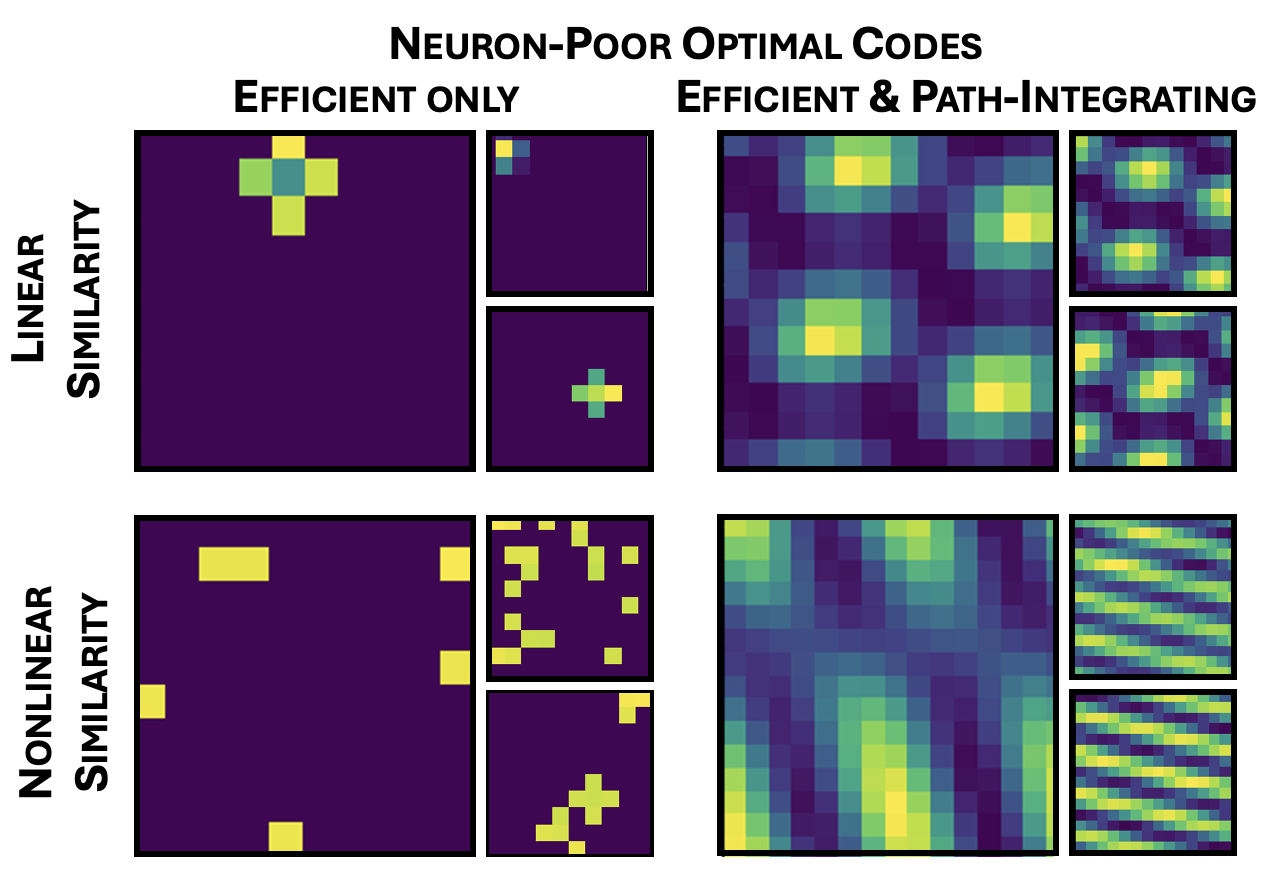}
    \caption{\textbf{A Space of Optimal Codes} We optimise a nonnegative, unit-norm representation of position to minimise a similarity matching objective either linear,~\cref{eq:sim_match}, or nonlinear,~\cref{eq:non_linear_similarity}, with or without a path-integrating constraint,~\cref{eq:actionability}. With more neurons than positions all choices lead to place cells (not shown). With few neurons and no path-integration (left column) we get place cells with a linear objective, and random multifields with a nonlinear objective (see also fig 15C, \citep{dorrell2023actionable}). Adding a path-integration constraint leads to either one grid module for the linear similarity loss, or multiple under the nonlinear loss (for more discussion, see \citep{dorrell2023actionable}).} 
    \label{fig:grid_lit-path_int_codes}
\end{figure}

When the number of neurons are scarce, under linear losses place cells are optimal without a path-integration requirement and a single module of grid cells when path-integration is required. Both these codes are \textit{not} combinatorial as linear losses do not profit from combinatorial codes, ~\cref{fig:grid_lit-path_int_codes} top. On the other hand, with a nonlinear loss multifield (combinatorial) place cells are optimal without a path-integration requirement, while multiple modules of axis-aligned grid cells are optimal when path-integration is required,~\cref{fig:grid_lit-path_int_codes} bottom. 

\subsection{Efficient Coding using Multimodular Codes}\label{sec:efficient_coding_multimodular}

We have discussed how combining low nonnegative firing rates with a sufficiently flexible nonlinear decoding and path-integration leads to multiple modules of translationally symmetric grid cells. We now consider one final normative question: how should these modules actually be structured? What lattice should they use (e.g. square or hexagon)? What should the relative size and orientation between modules be? And how many neurons per module?

The first forays in tackling these question assumed a multimodular structure and then optimised the remaining parameters to maximise the mutual information between neural activity and position, through proxies such as the Fisher information. Having demonstrated that a multimodular grid code encodes space with a higher accuracy than a place cell code \citep{sreenivasan2011grid,mathis2012optimal}, it was found that, of all lattice choices, hexagonal lattices were optimal \citep{mathis2012optimal,mathis2012resolution}. Subsequent related works derived similar results\citep{stemmler2015connecting,wei2015principle} and emphasised the effect of independent per-module noise \citep{towse2014optimal}. Further, the same set of ideas have been used to suggest that fewer neurons are required in grid modules with longer lengthscales \citep{mosheiff2017efficient}. 

Much work then analysed the optimal choice of ratio between the lattice lengthscales of successive grid modules. Early experimental work suggested a geometric progression of lengthscales with a constant ratio of between 1.4 and 1.7 \citep{stensola2012entorhinal,barry2007experience}, findings that were matched by multiple theoretical accounts \citep{wei2015principle,mathis2013multiscale}. However, it remains unclear whether a geometric progression model is actually well-matched to data, especially as measuring multiple modules simultaneously is technically difficult. Indeed, recent models based on developmental arguments predict non-geometric ratios that also appear to match measurements well \citep{khona2025global}, while our own work which suggests that grid modules should be related by non harmonic ratios \citep{dorrell2023actionable}. 

Grid modules are not only defined by their lengthscale, but also the relative orientation to other modules. To understand these relative orientations, we used the same efficient coding arguments (that show multiple modules of grid cells are optimal) to predict that successive grid modules should be oriented at small angles relative to one another \citep{dorrell2023actionable}, matching measurements \citep{stensola2012entorhinal,lykken2025functional}. Finally, encoding arguments have also proved useful at understanding how grid cells code 1D space \citep{rebecca2025spatial}, the alignment of grid axes to square rooms \citep{rebecca2025spatial}, and the changing of grid lattice parameters to different room shapes \citep{stensola2012entorhinal,dorrell2023actionable}. 

In sum, having arrived at a multimodular structure, efficient coding is a useful framework for understanding the details of the multimodular arrangement.

\section{Discussion}\label{sec:grid_lit_disc}

Over a decade of normative grid cell theorising points to a core claim: grid cells form a (1) high-fidelity, (2) path-integrating, (3) biologically-plausible code for space. In contrast, normative attempts to explain grid cells without path-integration cannot match their translational symmetry,~\cref{sec:grid_lit_efficient_coding}; and theories using `overly linear' measures of coding capacity struggle to explain multimodular structure,~\cref{sec:nonlinear_decoding}. This coheres with mechanistic and perturbative work to support a compelling narrative regarding the grid cell code. 

There remain puzzles. While models based on action dependent weight matrices recover the multi-modular axis-aligned structure of grid cells in multiple models \citep{dorrell2023actionable, schaeffer2023self}, these models are unable to model the conjunctive grid cells. Models using standard RNNs can make statements about precise velocity update mechanisms \citep{sorscher2023unified,schoyen2023coherently,chu2025unfolding}, but do so in ways that don't match biology \citep{schoyen2023coherently,chu2025unfolding}, are at times badly behaved \citep{schaeffer2022free,schoyen2023coherently,pettersen2024self}, and struggle to produce multiple modules of grid cells. As such, a normative model that cohesively captures all four grid cell phenomena we began with remains at large. That said, it seems likely that a careful combination of the best parts of existing models might succeed. We now discuss two broader open questions, and a few implications of this body of work.

\subsection{Future Work}


\paragraph{Grid Cells in Other Spaces} We have focused on grid cells in 2D, a natural question is how might they behave in other spaces. Normative theories of path-integrable representations naturally generalise to other spaces, and almost always predict multiple modules densely packed lattices in other spaces \citep{stemmler2015connecting,dorrell2023actionable}, matching similar formulations in one dimension \citep{aceituno2024theoretical}. However, it appears that grid cells are a bespoke 2-dimensional system: 1-dimensional maps are understood by mapping onto a slice of the grid lattice \citep{yoon2016grid,jacob2019path,rebecca2025spatial}; conversely, 3D grid cells appear to have multiple randomly scattered fields \citep{ginosar2021locally,grieves2021irregular}, in contrast to either the models discussed so far, and more boutique projection models \citep{klukas2020efficient}. Models have been proposed that cohesively capture some aspects of both 2D and 3D coding \citep{ginosar2021locally}, but, as reviewed,~\cref{app:dense_packing}, they do a poor job at fitting 2D behaviour. Whether there is some preserved structure in the 3D recordings, or a more general model that explains how grid cells encode spaces beyond 2D remains a topic for further work.


\paragraph{Warping of Grid Cells to Environments or Rewards} One finding is that grid cells don't always look so... griddy. In trapezoidal environments the lattice bends along the walls \citep{krupic2015grid}, the lattice lengthscale gets smaller near boundaries \citep{hagglund2019grid}, in large environments there are often inhomogeneities \citep{stensola2015shearing,gutierrez2025tiling} (though these sometimes disappear with experience \citep{carpenter2015grid}), grid fields warp in response to rewards \citep{boccara2019entorhinal}, and the grid metric stretches in inhomogeneous environments \citep{wen2024one}. Some models have taken this at face value, and attempted to normatively explain the warped grid responses, for example as the optimal code for uncertainty \citep{kang2023spatial}. Others have argued that the warping is the effect of an optimally mixed encoding of additional variables beyond space \citep{whittington2023disentanglement,dorrell2025range}. A final approach models these effects as a re-centering of the grid code in response to an external cue, such as a boundary \citep{ocko2018emergent}. Since these last two approaches understand inhomogoneities through perturbations to an underlying pure grid cell code, they are consistent with existing normative theories. Indeed, the observed rate maps could represent pure grid code after a spatially dependent recentering operation, making perfect grids appear bent in some environments or towards some rewards. However, the same is not true of the first model, and, as yet, no model is able to bridge these two domains clearly.


\subsection{Some Implications}

\paragraph{How constrained are these ideas?} Across this body of work, the way in which the three ideas: `high-fidelity', `path-integrable', or `biological', have been formalised has varied. This is a good thing, demonstrating robustness to ad hoc modelling choices. However, some recurring motifs stand-out. In all cases, the biological constraints limit the capacity of the system (e.g. by limiting the range of firing rates), and ensure the problem is not rotationally invariant, using a nonnegativity constraint either on neural firing or on weights. Similarly, path-integration always implies some mechanism for forward modelling: predicting the next encoding from your previous encoding and an action. Finally, the implementation of a high-fidelity code has relied on some form of `functional nonlinearity' in the decoding loss. 



\paragraph{Single Neurons are Pleasingly Constraining} 
Broadly, it is potentially unclear how much measuring a small number of single neurons can reliably guide our understanding of the brain \citep{whittington2026much}. Alternative approaches advocate for studying population-level metrics (e.g. \cite{stringer2019high}). There are only $\sim10000$s grid cells in a rat (using estimates from \citep{clark2024task,gatome2010number,diehl2017grid}), yet reviewing this literature we see that it has been incredibly constraining. Fitting just four high-level properties of the system has identified a core set of computational principles across models, and has proved adept at discounting alternative hypotheses. This is a ringing endorsement for the plodding progress of standard neuroscience.

\paragraph{RNNs as neural models} Using task-optimised neural networks as neural models is somewhat controversial; in complex tasks they are often as confusing as the brain \citep{banino2018vector}, limiting the insights we can gain from them. Yet the grid cell literature presents a compelling case for their power when coupled with clear experimentation, and thorough analysis. Task-optimised networks permit you to try a variety of hypotheses relatively quickly and flexibly. Their downside is that the signal you measure might have been caused by any number of choices made in architecture, training, or regularisation, and it is often hard to test for all of these. Simplifying the model to the point where theoretical work is possible can provide insight, allowing fine-tuning of the RNN experiments. For grid cells, iterations of this cycle seem to have nearly converged. We are optimists, and hope this will be more broadly true, suggesting a version of `analytic connectionism' that pairs careful theory and network modelling. Yet, we note that in the grid cell world this has already taken a decade of intense arguments: it is not necessarily easy.

\paragraph{The Power of Normative Modelling} Early work demonstrated that multimodular grid cells are a much more informative code for space than place cells \citep{mathis2012optimal}, leading to a view of grid cells as an efficient code for space. We hope this review has disabused you of this notion: grid cells are an efficient, but not the most efficient code for space---rather, they are the most efficient {\it path-integrating} code for space: random multifield place cells are the most efficient code,~\cref{fig:grid_lit-path_int_codes}. This highlights a role for normative modelling: by searching amongst all possible codes we are forced to consider all alternatives, highlighting how, if the only goal was efficiency, the best choice would never be grid cells. This null result cleanly highlights a key missing ingredient: path-integration.

\subsection{Conclusion}

In conclusion, the manifold structures present in the grid cell system have provided impressive constraints for normative theorising. After much work, the field has settled on a consistent set of normative theories: grid cells are a high-fidelity, path-integrable, biological (i.e. constrained and axis-dependent) code for space, agreeing with mechanistic and experimental work. In the future we hope these insights will generalise to grid cells in more complex settings, other neural systems, and provide broad lessons for successful normative theorising.

\paragraph{Code:} A simple jupyter notebook to generate the optimal representations in~\cref{fig:grid_lit-path_int_codes} and~\cref{fig:Pettersen_Largesig_Place} can be found at \url{https://github.com/WilburDoz/If_Grid_Cells_are_the_answer_what_is_the_Question.git}.

\paragraph{Acknowledgements} We thank Ben Sorscher, Mikhail Khona, Rylan Schaeffer, Tim Behrens, and Peter Doohan for reading earlier drafts of this work, and especially highlight Charles Burns and Markus Pettersen for their detailed and helpful comments.

We thank the following funding sources: Gatsby Charitable Foundation (GAT3755; W.D.); Sir Henry Wellcome Post-doctoral Fellowship (222817/Z/21/Z; J.C.R.W); European Research Council Starting Grant (NARFB/101222868; J.C.R.W).



\printbibliography

@inproceedings{whittington2023disentanglement,
  title={Disentanglement with biological constraints: A theory of functional cell types},
  author={Whittington, James CR and Dorrell, Will and Ganguli, Surya and Behrens, Timothy},
  booktitle={The Eleventh International Conference on Learning Representations},
  year={2023}
}

@inproceedings{xu2020theory,
    title={A Theory of Usable Information under Computational Constraints},
    author={Yilun Xu and Shengjia Zhao and Jiaming Song and Russell Stewart and Stefano Ermon},
    booktitle={International Conference on Learning Representations},
    year={2020},
}

@article{hafting2005microstructure,
  title={Microstructure of a spatial map in the entorhinal cortex},
  author={Hafting, Torkel and Fyhn, Marianne and Molden, Sturla and Moser, May-Britt and Moser, Edvard I},
  journal={Nature},
  volume={436},
  number={7052},
  pages={801--806},
  year={2005},
  publisher={Nature Publishing Group UK London}}

@article{whittington2020tolman,
  title={The Tolman-Eichenbaum machine: unifying space and relational memory through generalization in the hippocampal formation},
  author={Whittington, James CR and Muller, Timothy H and Mark, Shirley and Chen, Guifen and Barry, Caswell and Burgess, Neil and Behrens, Timothy EJ},
  journal={Cell},
  volume={183},
  number={5},
  pages={1249--1263},
year={2020},
  publisher={Elsevier}
}

@article{logiaco2021thalamic,
  title={Thalamic control of cortical dynamics in a model of flexible motor sequencing},
  author={Logiaco, Laureline and Abbott, LF and Escola, Sean},
  journal={Cell reports},
  volume={35},
  number={9},
  year={2021},
  publisher={Elsevier}
}

@article{stensola2012entorhinal,
  title={The entorhinal grid map is discretized},
  author={Stensola, Hanne and Stensola, Tor and Solstad, Trygve and Fr{\o}land, Kristian and Moser, May-Britt and Moser, Edvard I},
  journal={Nature},
  volume={492},
  number={7427},
  pages={72--78},
  year={2012},
  publisher={Nature Publishing Group UK London}
}

@inproceedings{dorrell2025range,
  title={Range, not Independence, Drives Modularity in Biologically Inspired Representations},
  author={Dorrell, William and Hsu, Kyle and Hollingsworth, Luke and Lee, Jin Hwa and Wu, Jiajun and Finn, Chelsea and Latham, Peter E and Behrens, Timothy Edward John and Whittington, James CR},
  booktitle={The Thirteenth International Conference on Learning Representations},
year = {2025},
}

@inproceedings{
dorrell2023actionable,
title={Actionable Neural Representations: Grid Cells from Minimal Constraints},
author={Dorrell, Will and Latham, Peter and Behrens, Timothy and Whittington, James C. R.},
booktitle={The Eleventh International Conference on Learning Representations },
year={2023},
url={https://openreview.net/forum?id=xfqDe72zh41}
}

@article{stensola2015shearing,
  title={Shearing-induced asymmetry in entorhinal grid cells},
  author={Stensola, Tor and Stensola, Hanne and Moser, May-Britt and Moser, Edvard I},
  journal={Nature},
  volume={518},
  number={7538},
  pages={207--212},
  year={2015},
  publisher={Nature Publishing Group}
}

@article{gao2021path,
  title={On Path Integration of grid cells: isotropic metric, conformal embedding and group representation},
  author={Gao, Ruiqi and Xie, Jianwen and Wei, Xue-Xin and Zhu, Song-Chun and Wu, Ying Nian},
  journal={Advances in neural information processing systems},
  volume={34},
  year={2021}
}

@article{mathis2012optimal,
  title={Optimal population codes for space: grid cells outperform place cells},
  author={Mathis, Alexander and Herz, Andreas VM and Stemmler, Martin},
  journal={Neural computation},
  volume={24},
  number={9},
  pages={2280--2317},
  year={2012},
  publisher={MIT Press One Rogers Street, Cambridge, MA 02142-1209, USA journals-info~…}
}

@article{stemmler2015connecting,
  title={Connecting multiple spatial scales to decode the population activity of grid cells},
  author={Stemmler, Martin and Mathis, Alexander and Herz, Andreas VM},
  journal={Science Advances},
  volume={1},
  number={11},
  pages={e1500816},
  year={2015},
  publisher={American Association for the Advancement of Science}
}

@article{mathis2012resolution,
  title={Resolution of nested neuronal representations can be exponential in the number of neurons},
  author={Mathis, Alexander and Herz, Andreas VM and Stemmler, Martin B},
  journal={Physical review letters},
  volume={109},
  number={1},
  pages={018103},
  year={2012},
  publisher={APS}
}

@article{sreenivasan2011grid,
  title={Grid cells generate an analog error-correcting code for singularly precise neural computation},
  author={Sreenivasan, Sameet and Fiete, Ila},
  journal={Nature neuroscience},
  volume={14},
  number={10},
  pages={1330--1337},
  year={2011},
  publisher={Nature Publishing Group}
}

@article{wei2015principle,
  title={A principle of economy predicts the functional architecture of grid cells},
  author={Wei, Xue-Xin and Prentice, Jason and Balasubramanian, Vijay},
  journal={Elife},
  volume={4},
  pages={e08362},
  year={2015},
  publisher={eLife Sciences Publications Limited}
}

@article{cueva2018emergence,
  title={Emergence of grid-like representations by training recurrent neural networks to perform spatial localization},
  author={Cueva, Christopher J and Wei, Xue-Xin},
  journal={arXiv preprint arXiv:1803.07770},
  year={2018}
}

@article{sorscher2019unified,
  title={A unified theory for the origin of grid cells through the lens of pattern formation},
  author={Sorscher, Ben and Mel, Gabriel and Ganguli, Surya and Ocko, Samuel},
  journal={Advances in neural information processing systems},
  volume={32},
  year={2019}
}

@article{dordek2016extracting,
  title={Extracting grid cell characteristics from place cell inputs using non-negative principal component analysis},
  author={Dordek, Yedidyah and Soudry, Daniel and Meir, Ron and Derdikman, Dori},
  journal={Elife},
  volume={5},
  pages={e10094},
  year={2016},
  publisher={eLife Sciences Publications Limited}
}

@article{stachenfeld2017hippocampus,
  title={The hippocampus as a predictive map},
  author={Stachenfeld, Kimberly L and Botvinick, Matthew M and Gershman, Samuel J},
  journal={Nature neuroscience},
  volume={20},
  number={11},
  pages={1643--1653},
  year={2017},
  publisher={Nature Publishing Group}
}

@article{ginosar2021locally,
  title={Locally ordered representation of 3D space in the entorhinal cortex},
  author={Ginosar, Gily and Aljadeff, Johnatan and Burak, Yoram and Sompolinsky, Haim and Las, Liora and Ulanovsky, Nachum},
  journal={Nature},
  volume={596},
  number={7872},
  pages={404--409},
  year={2021},
  publisher={Nature Publishing Group}
}

@article{grieves2021irregular,
  title={Irregular distribution of grid cell firing fields in rats exploring a 3D volumetric space},
  author={Grieves, Roddy M and Jedidi-Ayoub, Selim and Mishchanchuk, Karyna and Liu, Anyi and Renaudineau, Sophie and Duvelle, {\'E}l{\'e}onore and Jeffery, Kate J},
  journal={Nature neuroscience},
  volume={24},
  number={11},
  pages={1567--1573},
  year={2021},
  publisher={Nature Publishing Group}
}

@article{burak2009accurate,
  title={Accurate path integration in continuous attractor network models of grid cells},
  author={Burak, Yoram and Fiete, Ila R},
  journal={PLoS computational biology},
  volume={5},
  number={2},
  pages={e1000291},
  year={2009},
  publisher={Public Library of Science San Francisco, USA}
}

@article{sengupta2018manifold,
  title={Manifold-tiling localized receptive fields are optimal in similarity-preserving neural networks},
  author={Sengupta, Anirvan and Pehlevan, Cengiz and Tepper, Mariano and Genkin, Alexander and Chklovskii, Dmitri},
  journal={Advances in neural information processing systems},
  volume={31},
  year={2018}
}

@article{boccara2019entorhinal,
  title={The entorhinal cognitive map is attracted to goals},
  author={Boccara, Charlotte N and Nardin, Michele and Stella, Federico and O’Neill, Joseph and Csicsvari, Jozsef},
  journal={Science},
  volume={363},
  number={6434},
  pages={1443--1447},
  year={2019},
  publisher={American Association for the Advancement of Science}
}

@article{krupic2015grid,
  title={Grid cell symmetry is shaped by environmental geometry},
  author={Krupic, Julija and Bauza, Marius and Burton, Stephen and Barry, Caswell and O’Keefe, John},
  journal={Nature},
  volume={518},
  number={7538},
  pages={232--235},
  year={2015},
  publisher={Nature Publishing Group}
}

@article{issa2012universal,
  title={Universal conditions for exact path integration in neural systems},
  author={Issa, John B and Zhang, Kechen},
  journal={Proceedings of the National Academy of Sciences},
  volume={109},
  number={17},
  pages={6716--6720},
  year={2012},
  publisher={National Acad Sciences}
}

@article{schaeffer2022free,
    title = {No Free Lunch from Deep Learning in Neuroscience: A Case Study through Models of the Entorhinal-Hippocampal Circuit},
    author={Schaeffer, Rylan and Khona, Mikail and Fiete, Ila},
    journal={ICML 2022 Workshop AI4Science},
    year={2022}
}

@article{tolman1948cognitive,
  title={Cognitive maps in rats and men.},
  author={Tolman, Edward C},
  journal={Psychological review},
  volume={55},
  number={4},
  year={1948}
}

@article{laughlin1981simple,
  title={A simple coding procedure enhances a neuron's information capacity},
  author={Laughlin, Simon},
  journal={Zeitschrift f{\"u}r Naturforschung c},
  volume={36},
  pages={910--912},
  year={1981},
  publisher={De Gruyter}
}

@article{olshausen1996emergence,
  title={Emergence of simple-cell receptive field properties by learning a sparse code for natural images},
  author={Olshausen, Bruno A and Field, David J},
  journal={Nature},
  volume={381},
  number={6583},
  pages={607--609},
  year={1996},
  publisher={Nature Publishing Group}
}

@article{ocko2018emergent,
  title={Emergent elasticity in the neural code for space},
  author={Ocko, Samuel A and Hardcastle, Kiah and Giocomo, Lisa M and Ganguli, Surya},
  journal={Proceedings of the National Academy of Sciences},
  volume={115},
  number={50},
  pages={E11798--E11806},
  year={2018},
  publisher={National Acad Sciences}
}

@article{kim2017ring,
  title={Ring attractor dynamics in the Drosophila central brain},
  author={Kim, Sung Soo and Rouault, Herv{\'e} and Druckmann, Shaul and Jayaraman, Vivek},
  journal={Science},
  volume={356},
  number={6340},
  pages={849--853},
  year={2017},
  publisher={American Association for the Advancement of Science}
}

@article{yu2020prediction,
  title={Prediction and Generalisation over Directed Actions by Grid Cells},
  author={Yu, Changmin and Behrens, Timothy EJ and Burgess, Neil},
  journal={arXiv preprint arXiv:2006.03355},
  year={2020}
}

@article{gao2019learning,
  title={Learning grid cells as vector representation of self-position coupled with matrix representation of self-motion},
  author={Gao, Ruiqi and Xie, Jianwen and Zhu, Song-Chun and Wu, Ying Nian},
  journal={Internal Conference on Learning Representations},
  year={2019}
}

@article{sorscher2022and,
  title={When and why grid cells appear or not in trained path integrators},
  author={Sorscher, Ben and Mel, Gabriel C and Nayebi, Aran and Giocomo, Lisa and Yamins, Daniel and Ganguli, Surya},
  journal={bioRxiv},
  year={2022},
  publisher={Cold Spring Harbor Laboratory}
}

@article{kang2023spatial,
  title={Spatial uncertainty and environmental geometry in navigation},
  author={Kang, Yul HR and Wolpert, Daniel M and Lengyel, M{\'a}t{\'e}},
  journal={bioRxiv},
  pages={2023--01},
  year={2023},
  publisher={Cold Spring Harbor Laboratory}
}

@article{banino2018vector,
  title={Vector-based navigation using grid-like representations in artificial agents},
  author={Banino, Andrea and Barry, Caswell and Uria, Benigno and Blundell, Charles and Lillicrap, Timothy and Mirowski, Piotr and Pritzel, Alexander and Chadwick, Martin J and Degris, Thomas and Modayil, Joseph and others},
  journal={Nature},
  volume={557},
  number={7705},
  pages={429--433},
  year={2018},
  publisher={Nature Publishing Group}
}

@article{vollan2025left,
  title={Left--right-alternating theta sweeps in entorhinal--hippocampal maps of space},
  author={Vollan, Abraham Z and Gardner, Richard J and Moser, May-Britt and Moser, Edvard I},
  journal={Nature},
  pages={1--11},
  year={2025},
  publisher={Nature Publishing Group UK London}
}

@article{sargolini2006conjunctive,
  title={Conjunctive representation of position, direction, and velocity in entorhinal cortex},
  author={Sargolini, Francesca and Fyhn, Marianne and Hafting, Torkel and McNaughton, Bruce L and Witter, Menno P and Moser, May-Britt and Moser, Edvard I},
  journal={Science},
  volume={312},
  number={5774},
  pages={758--762},
  year={2006},
  publisher={American Association for the Advancement of Science}
}

@article{mcnaughton2006path,
  title={Path integration and the neural basis of the'cognitive map'},
  author={McNaughton, Bruce L and Battaglia, Francesco P and Jensen, Ole and Moser, Edvard I and Moser, May-Britt},
  journal={Nature Reviews Neuroscience},
  volume={7},
  number={8},
  pages={663--678},
  year={2006},
  publisher={Nature Publishing Group UK London}
}

@article{skaggs1994model,
  title={A model of the neural basis of the rat's sense of direction},
  author={Skaggs, William and Knierim, James and Kudrimoti, Hemant and McNaughton, Bruce},
  journal={Advances in neural information processing systems},
  volume={7},
  year={1994}
}

@article{redish1996coupled,
  title={A coupled attractor model of the rodent head direction system},
  author={Redish, A David and Elga, Adam N and Touretzky, David S},
  journal={Network: computation in neural systems},
  volume={7},
  number={4},
  pages={671},
  year={1996},
  publisher={IOP Publishing}
}

@article{touretzky1996theory,
  title={Theory of rodent navigation based on interacting representations of space},
  author={Touretzky, David S and Redish, A David},
  journal={Hippocampus},
  volume={6},
  number={3},
  pages={247--270},
  year={1996},
  publisher={Wiley Online Library}
}

@article{conklin2005controlled,
  title={A controlled attractor network model of path integration in the rat},
  author={Conklin, John and Eliasmith, Chris},
  journal={Journal of computational neuroscience},
  volume={18},
  pages={183--203},
  year={2005},
  publisher={Springer}
}

@article{samsonovich1997path,
  title={Path integration and cognitive mapping in a continuous attractor neural network model},
  author={Samsonovich, Alexei and McNaughton, Bruce L},
  journal={Journal of Neuroscience},
  volume={17},
  number={15},
  pages={5900--5920},
  year={1997},
  publisher={Society for Neuroscience}
}

@article{khona2025global,
  title={Global modules robustly emerge from local interactions and smooth gradients},
  author={Khona, Mikail and Chandra, Sarthak and Fiete, Ila},
  journal={Nature},
  pages={1--10},
  year={2025},
  publisher={Nature Publishing Group UK London}
}

@article{kang2019geometric,
  title={A geometric attractor mechanism for self-organization of entorhinal grid modules},
  author={Kang, Louis and Balasubramanian, Vijay},
  journal={Elife},
  volume={8},
  pages={e46687},
  year={2019},
  publisher={eLife Sciences Publications, Ltd}
}

@article{ying2022disruption,
  title={Disruption of the grid cell network in a mouse model of early Alzheimer’s disease},
  author={Ying, Johnson and Keinath, Alexandra T and Lavoie, Raphael and Vigneault, Erika and El Mestikawy, Salah and Brandon, Mark P},
  journal={Nature Communications},
  volume={13},
  number={1},
  pages={886},
  year={2022},
  publisher={Nature Publishing Group UK London}
}

@article{gil2018impaired,
  title={Impaired path integration in mice with disrupted grid cell firing},
  author={Gil, Mariana and Ancau, Mihai and Schlesiger, Magdalene I and Neitz, Angela and Allen, Kevin and De Marco, Rodrigo J and Monyer, Hannah},
  journal={Nature neuroscience},
  volume={21},
  number={1},
  pages={81--91},
  year={2018},
  publisher={Nature Publishing Group US New York}
}

@article{jun2020disrupted,
  title={Disrupted place cell remapping and impaired grid cells in a knockin model of Alzheimer's disease},
  author={Jun, Heechul and Bramian, Allen and Soma, Shogo and Saito, Takashi and Saido, Takaomi C and Igarashi, Kei M},
  journal={Neuron},
  volume={107},
  number={6},
  pages={1095--1112},
  year={2020},
  publisher={Elsevier}
}

@article{kunz2015reduced,
  title={Reduced grid-cell--like representations in adults at genetic risk for Alzheimer’s disease},
  author={Kunz, Lukas and Schr{\"o}der, Tobias Navarro and Lee, Hweeling and Montag, Christian and Lachmann, Bernd and Sariyska, Rayna and Reuter, Martin and Stirnberg, R{\"u}diger and St{\"o}cker, Tony and Messing-Floeter, Paul Christian and others},
  journal={Science},
  volume={350},
  number={6259},
  pages={430--433},
  year={2015},
  publisher={American Association for the Advancement of Science}
}

@article{steffenach2005spatial,
  title={Spatial memory in the rat requires the dorsolateral band of the entorhinal cortex},
  author={Steffenach, Hill-Aina and Witter, Menno and Moser, May-Britt and Moser, Edvard I},
  journal={Neuron},
  volume={45},
  number={2},
  pages={301--313},
  year={2005},
  publisher={Elsevier}
}

@article{van2013distinct,
  title={Distinct roles of medial and lateral entorhinal cortex in spatial cognition},
  author={Van Cauter, Tiffany and Camon, Jeremy and Alvernhe, Alice and Elduayen, Coralie and Sargolini, Francesca and Save, Etienne},
  journal={Cerebral Cortex},
  volume={23},
  number={2},
  pages={451--459},
  year={2013},
  publisher={Oxford University Press}
}

@article{mok2019non,
  title={A non-spatial account of place and grid cells based on clustering models of concept learning},
  author={Mok, Robert M and Love, Bradley C},
  journal={Nature communications},
  volume={10},
  number={1},
  pages={5685},
  year={2019},
  publisher={Nature Publishing Group UK London}
}

@article{sorscher2023unified,
  title={A unified theory for the computational and mechanistic origins of grid cells},
  author={Sorscher, Ben and Mel, Gabriel C and Ocko, Samuel A and Giocomo, Lisa M and Ganguli, Surya},
  journal={Neuron},
  volume={111},
  number={1},
  pages={121--137},
  year={2023},
  publisher={Elsevier}
}

@article{schoyen2023coherently,
  title={Coherently remapping toroidal cells but not grid cells are responsible for path integration in virtual agents},
  author={Sch{\o}yen, Vemund and Pettersen, Markus Borud and Holzhausen, Konstantin and Fyhn, Marianne and Malthe-S{\o}renssen, Anders and Lepper{\o}d, Mikkel Elle},
  journal={Iscience},
  volume={26},
  number={11},
  year={2023},
  publisher={Elsevier}
}

@article{tang2024learning,
  title={Learning grid cells by predictive coding},
  author={Tang, Mufeng and Barron, Helen and Bogacz, Rafal},
  journal={arXiv preprint arXiv:2410.01022},
  year={2024}
}

@article{schaeffer2023testing,
  title={Testing assumptions underlying a unified theory for the origin of grid cells},
  author={Schaeffer, Rylan and Khona, Mikail and Bertagnoli, Adrian and Koyejo, Sanmi and Fiete, Ila Rani},
  journal={arXiv preprint arXiv:2311.16295},
  year={2023}
}

@article{schaeffer2023disentangling,
  title={Disentangling Fact from Grid Cell Fiction in Trained Deep Path Integrators},
  author={Schaeffer, Rylan and Khona, Mikail and Koyejo, Sanmi and Fiete, Ila Rani},
  journal={ArXiv},
  pages={arXiv--2312},
  year={2023}
}

@article{deighton2024higher,
  title={Higher-Order Spatial Information for Self-Supervised Place Cell Learning},
  author={Deighton, Jared and Mackey, Wyatt and Schizas, Ioannis and Boothe Jr, David L and Maroulas, Vasileios},
  journal={arXiv preprint arXiv:2407.06195},
  year={2024}
}

@article{rich2014large,
  title={Large environments reveal the statistical structure governing hippocampal representations},
  author={Rich, P Dylan and Liaw, Hua-Peng and Lee, Albert K},
  journal={Science},
  volume={345},
  number={6198},
  pages={814--817},
  year={2014},
  publisher={American Association for the Advancement of Science}
}

@article{eliav2021multiscale,
  title={Multiscale representation of very large environments in the hippocampus of flying bats},
  author={Eliav, Tamir and Maimon, Shir R and Aljadeff, Johnatan and Tsodyks, Misha and Ginosar, Gily and Las, Liora and Ulanovsky, Nachum},
  journal={Science},
  volume={372},
  number={6545},
  pages={eabg4020},
  year={2021},
  publisher={American Association for the Advancement of Science}
}

@article{burgess2007oscillatory,
  title={An oscillatory interference model of grid cell firing},
  author={Burgess, Neil and Barry, Caswell and O'keefe, John},
  journal={Hippocampus},
  volume={17},
  number={9},
  pages={801--812},
  year={2007},
  publisher={Wiley Online Library}
}

@article{huber2025memory,
  title={A memory model of rodent spatial navigation in which place cells are memories arranged in a grid and grid cells are non-spatial},
  author={Huber, David E},
  journal={eLife},
  volume={13},
  pages={RP95733},
  year={2025},
  publisher={eLife Sciences Publications Limited}
}

@inproceedings{xu2025conformal,
  title={On conformal isometry of grid cells: Learning distance-preserving position embedding},
  author={Xu, Dehong and Gao, Ruiqi and Zhang, Wenhao and Wei, Xue-Xin and Wu, Ying Nian},
  booktitle={The Thirteenth International Conference on Learning Representations},
  year={2025}
}

@article{pettersen2024self,
  title={Self-supervised grid cells without path integration},
  author={Pettersen, Markus and Sch{\o}yen, Vemund Sigmundson and {\O}stby, Mattis Dals{\ae}tra and Malthe-S{\o}renssen, Anders and Lepper{\o}d, Mikkel Elle},
  journal={bioRxiv},
  pages={2024--05},
  year={2024},
  publisher={Cold Spring Harbor Laboratory}
}

@article{towse2014optimal,
  title={Optimal configurations of spatial scale for grid cell firing under noise and uncertainty},
  author={Towse, Benjamin W and Barry, Caswell and Bush, Daniel and Burgess, Neil},
  journal={Philosophical Transactions of the Royal Society B: Biological Sciences},
  volume={369},
  number={1635},
  pages={20130290},
  year={2014},
  publisher={The Royal Society}
}

@phdthesis{ma2020towards,
  title={Towards a theory for the emergence of grid and place cell codes},
  author={Ma, Tzuhsuan},
  year={2020},
  school={Massachusetts Institute of Technology}
}

@article{mathis2013multiscale,
  title={Multiscale codes in the nervous system: the problem of noise correlations and the ambiguity of periodic scales},
  author={Mathis, Alexander and Herz, Andreas VM and Stemmler, Martin B},
  journal={Physical Review E—Statistical, Nonlinear, and Soft Matter Physics},
  volume={88},
  number={2},
  pages={022713},
  year={2013},
  publisher={APS}
}

@article{kubie2012linear,
  title={Linear look-ahead in conjunctive cells: an entorhinal mechanism for vector-based navigation},
  author={Kubie, John L and Fenton, Andr{\'e} A},
  journal={Frontiers in neural circuits},
  volume={6},
  pages={20},
  year={2012},
  publisher={Frontiers Research Foundation}
}

@article{rebecca2025spatial,
  title={Spatial periodicity in grid cell firing is explained by a neural sequence code of 2-D trajectories},
  author={Rebecca, RG and Ascoli, Giorgio A and Sutton, Nate M and Dannenberg, Holger},
  journal={eLife},
  volume={13},
  pages={RP96627},
  year={2025},
  publisher={eLife Sciences Publications Limited}
}

@article{whittington2018generalisation,
  title={Generalisation of structural knowledge in the hippocampal-entorhinal system},
  author={Whittington, James and Muller, Timothy and Mark, Shirely and Barry, Caswell and Behrens, Tim},
  journal={Advances in neural information processing systems},
  volume={31},
  year={2018}
}

@article{schaeffer2023self,
  title={Self-supervised learning of representations for space generates multi-modular grid cells},
  author={Schaeffer, Rylan and Khona, Mikail and Ma, Tzuhsuan and Eyzaguirre, Cristobal and Koyejo, Sanmi and Fiete, Ila},
  journal={Advances in Neural Information Processing Systems},
  volume={36},
  pages={23140--23157},
  year={2023}
}

@article{schoyen2025hexagons,
  title={Hexagons all the way down: Grid cells as a conformal isometric map of space},
  author={Sch{\o}yen, Vemund Sigmundson and Beshkov, Kosio and Pettersen, Markus Borud and Hermansen, Erik and Holzhausen, Konstantin and Malthe-S{\o}renssen, Anders and Fyhn, Marianne and Lepper{\o}d, Mikkel Elle},
  journal={PLOS Computational Biology},
  volume={21},
  number={2},
  pages={e1012804},
  year={2025},
  publisher={Public Library of Science San Francisco, CA USA}
}

@article{aceituno2024theoretical,
  title={Theoretical principles explain the structure of the insect head direction circuit},
  author={Aceituno, Pau Vilimelis and Dall'Osto, Dominic and Pisokas, Ioannis},
  journal={Elife},
  volume={13},
  pages={e91533},
  year={2024},
  publisher={eLife Sciences Publications Limited}
}

@article{yoon2016grid,
  title={Grid cell responses in 1D environments assessed as slices through a 2D lattice},
  author={Yoon, KiJung and Lewallen, Sam and Kinkhabwala, Amina A and Tank, David W and Fiete, Ila R},
  journal={Neuron},
  volume={89},
  number={5},
  pages={1086--1099},
  year={2016},
  publisher={Elsevier}
}

@article{jacob2019path,
  title={Path integration maintains spatial periodicity of grid cell firing in a 1D circular track},
  author={Jacob, Pierre-Yves and Capitano, Fabrizio and Poucet, Bruno and Save, Etienne and Sargolini, Francesca},
  journal={Nature communications},
  volume={10},
  number={1},
  pages={840},
  year={2019},
  publisher={Nature Publishing Group UK London}
}

@article{klukas2020efficient,
  title={Efficient and flexible representation of higher-dimensional cognitive variables with grid cells},
  author={Klukas, Mirko and Lewis, Marcus and Fiete, Ila},
  journal={PLoS computational biology},
  volume={16},
  number={4},
  pages={e1007796},
  year={2020},
  publisher={Public Library of Science San Francisco, CA USA}
}

@article{hagglund2019grid,
  title={Grid-cell distortion along geometric borders},
  author={H{\"a}gglund, Martin and M{\o}rreaunet, Maria and Moser, May-Britt and Moser, Edvard I},
  journal={Current Biology},
  volume={29},
  number={6},
  pages={1047--1054},
  year={2019},
  publisher={Elsevier}
}

@article{wen2024one,
  title={One-shot entorhinal maps enable flexible navigation in novel environments},
  author={Wen, John H and Sorscher, Ben and Aery Jones, Emily A and Ganguli, Surya and Giocomo, Lisa M},
  journal={Nature},
  volume={635},
  number={8040},
  pages={943--950},
  year={2024},
  publisher={Nature Publishing Group UK London}
}

@article{gutierrez2025tiling,
  title={Tiling of large-scaled environments by grid cells requires experience},
  author={Guti{\'e}rrez-Guzm{\'a}n, Blanca E and Hern{\'a}ndez-P{\'e}rez, J Jes{\'u}s and Dannenberg, Holger},
  journal={bioRxiv},
  year={2025}
}

@article{carpenter2015grid,
  title={Grid cells form a global representation of connected environments},
  author={Carpenter, Francis and Manson, Daniel and Jeffery, Kate and Burgess, Neil and Barry, Caswell},
  journal={Current Biology},
  volume={25},
  number={9},
  pages={1176--1182},
  year={2015},
  publisher={Elsevier}
}

@article{burgess2008grid,
  title={Grid cells and theta as oscillatory interference: theory and predictions},
  author={Burgess, Neil},
  journal={Hippocampus},
  volume={18},
  number={12},
  pages={1157--1174},
  year={2008},
  publisher={Wiley Online Library}
}

@article{hafting2008hippocampus,
  title={Hippocampus-independent phase precession in entorhinal grid cells},
  author={Hafting, Torkel and Fyhn, Marianne and Bonnevie, Tora and Moser, May-Britt and Moser, Edvard I},
  journal={Nature},
  volume={453},
  number={7199},
  pages={1248--1252},
  year={2008},
  publisher={Nature Publishing Group UK London}
}

@article{hulse2020mechanisms,
  title={Mechanisms underlying the neural computation of head direction},
  author={Hulse, Brad K and Jayaraman, Vivek},
  journal={Annual review of neuroscience},
  volume={43},
  number={1},
  pages={31--54},
  year={2020},
  publisher={Annual Reviews}
}

@article{waniek2020transition,
  title={Transition scale-spaces: A computational theory for the discretized entorhinal cortex},
  author={Waniek, Nicolai},
  journal={Neural computation},
  volume={32},
  number={2},
  pages={330--394},
  year={2020},
  publisher={MIT Press One Rogers Street, Cambridge, MA 02142-1209, USA journals-info~…}
}

@article{mosheiff2017efficient,
  title={An efficient coding theory for a dynamic trajectory predicts non-uniform allocation of entorhinal grid cells to modules},
  author={Mosheiff, Noga and Agmon, Haggai and Moriel, Avraham and Burak, Yoram},
  journal={PLoS computational biology},
  volume={13},
  number={6},
  pages={e1005597},
  year={2017},
  publisher={Public Library of Science San Francisco, CA USA}
}

@article{atick1990towards,
  title={Towards a theory of early visual processing},
  author={Atick, Joseph J and Redlich, A Norman},
  journal={Neural computation},
  volume={2},
  number={3},
  pages={308--320},
  year={1990},
  publisher={MIT Press}
}

@article{harland2021dorsal,
  title={Dorsal CA1 hippocampal place cells form a multi-scale representation of megaspace},
  author={Harland, Bruce and Contreras, Marco and Souder, Madeline and Fellous, Jean-Marc},
  journal={Current Biology},
  volume={31},
  number={10},
  pages={2178--2190},
  year={2021},
  publisher={Elsevier}
}

@article{lykken2025functional,
  title={Functional independence of entorhinal grid cell modules enables remapping in hippocampal place cells},
  author={Lykken, Christine M and Kanter, Benjamin R and Nagelhus, Anne and Carpenter, Jordan and Guardamagna, Matteo and Moser, Edvard I and Moser, May-Britt},
  journal={bioRxiv},
  pages={2025--09},
  year={2025},
  publisher={Cold Spring Harbor Laboratory}
}

@article{chu2025unfolding,
  title={Unfolding the Black Box of Recurrent Neural Networks for Path Integration},
  author={Chu, Tianhao and Wu, Yuling and Burgess, Neil and Ji, Zilong and Wu, Si},
  journal={bioRxiv},
  pages={2025--10},
  year={2025},
  publisher={Cold Spring Harbor Laboratory}
}

@article{bush2014hybrid,
  title={A hybrid oscillatory interference/continuous attractor network model of grid cell firing},
  author={Bush, Daniel and Burgess, Neil},
  journal={Journal of Neuroscience},
  volume={34},
  number={14},
  pages={5065--5079},
  year={2014},
  publisher={Society for Neuroscience}
}

@article{giocomo2008computation,
  title={Computation by oscillations: implications of experimental data for theoretical models of grid cells},
  author={Giocomo, Lisa M and Hasselmo, Michael E},
  journal={Hippocampus},
  volume={18},
  number={12},
  pages={1186--1199},
  year={2008},
  publisher={Wiley Online Library}
}

@article{hasselmo2008grid,
  title={Grid cell mechanisms and function: contributions of entorhinal persistent spiking and phase resetting},
  author={Hasselmo, Michael E},
  journal={Hippocampus},
  volume={18},
  number={12},
  pages={1213--1229},
  year={2008},
  publisher={Wiley Online Library}
}

@article{fuhs2006spin,
  title={A spin glass model of path integration in rat medial entorhinal cortex},
  author={Fuhs, Mark C and Touretzky, David S},
  journal={Journal of Neuroscience},
  volume={26},
  number={16},
  pages={4266--4276},
  year={2006},
  publisher={Society for Neuroscience}
}

@article{guanella2007model,
  title={A model of grid cells based on a twisted torus topology},
  author={Guanella, Alexis and Kiper, Daniel and Verschure, Paul},
  journal={International journal of neural systems},
  volume={17},
  number={04},
  pages={231--240},
  year={2007},
  publisher={World Scientific}
}

@article{pastoll2013feedback,
  title={Feedback inhibition enables theta-nested gamma oscillations and grid firing fields},
  author={Pastoll, Hugh and Solanka, Lukas and van Rossum, Mark CW and Nolan, Matthew F},
  journal={Neuron},
  volume={77},
  number={1},
  pages={141--154},
  year={2013},
  publisher={Elsevier}
}

@article{reifenstein2012grid,
  title={Grid cells in rat entorhinal cortex encode physical space with independent firing fields and phase precession at the single-trial level},
  author={Reifenstein, Eric T and Kempter, Richard and Schreiber, Susanne and Stemmler, Martin B and Herz, Andreas VM},
  journal={Proceedings of the National Academy of Sciences},
  volume={109},
  number={16},
  pages={6301--6306},
  year={2012},
  publisher={National Academy of Sciences}
}

@article{stringer2019high,
  title={High-dimensional geometry of population responses in visual cortex},
  author={Stringer, Carsen and Pachitariu, Marius and Steinmetz, Nicholas and Carandini, Matteo and Harris, Kenneth D},
  journal={Nature},
  volume={571},
  number={7765},
  pages={361--365},
  year={2019},
  publisher={Nature Publishing Group UK London}
}

@article{diehl2017grid,
  title={Grid and nongrid cells in medial entorhinal cortex represent spatial location and environmental features with complementary coding schemes},
  author={Diehl, Geoffrey W and Hon, Olivia J and Leutgeb, Stefan and Leutgeb, Jill K},
  journal={Neuron},
  volume={94},
  number={1},
  pages={83--92},
  year={2017},
  publisher={Elsevier}
}

@article{gatome2010number,
  title={Number estimates of neuronal phenotypes in layer II of the medial entorhinal cortex of rat and mouse},
  author={Gatome, CW and Slomianka, L and Lipp, HP and Amrein, I},
  journal={Neuroscience},
  volume={170},
  number={1},
  pages={156--165},
  year={2010},
  publisher={Elsevier}
}

@article{clark2024task,
  title={Task-anchored grid cell firing is selectively associated with successful path integration-dependent behaviour},
  author={Clark, Harry and Nolan, Matthew F},
  journal={Elife},
  volume={12},
  pages={RP89356},
  year={2024},
  publisher={eLife Sciences Publications Limited}
}

@article{barry2007experience,
  title={Experience-dependent rescaling of entorhinal grids},
  author={Barry, Caswell and Hayman, Robin and Burgess, Neil and Jeffery, Kathryn J},
  journal={Nature neuroscience},
  volume={10},
  number={6},
  pages={682--684},
  year={2007},
  publisher={Nature Publishing Group US New York}
}

@article{whittington2026much,
  title={How much neuroscience does a neuroscientist need to know?},
  author={Whittington, James CR and Dorrell, William},
  journal={arXiv preprint arXiv:2601.02063},
  year={2026}
}

\newpage

\appendix

\section{Hexagonal Lattices via Dense Packing Arguments}\label{app:dense_packing}

Hexagonal lattices are the densest packing of spheres in 2D space, or analogously, the best arrangement of sensors to minimise the average distance between all points in 2D space and the nearest sensor. One family of efficient-coding-only approaches use this idea to produce hexagonally tuned cells.

\cite{mok2019non} argue that place cells form a conceptual clustering of inputs: which place cells is active for each input corresponds to its cluster and the quality of the encoding is given by the resolution of the clustering\footnote{i.e. The best clustering would give every input its own cluster, the worst would assign all inputs to one cluster.}. They argue that space can be thought of as a uniform continuum of inputs to be explained, and that, thanks to dense packing, the optimal choice of a finite set of place cells (clusters) is a hexagonal grid. They then argue that grid cells are a measure of proximity between points in space and their nearest cluster---which in this model is a measure of how well fit that point is by the learnt clusters. Since the data is best explained at cluster centres this forms a hexagonal lattice. 

\cite{ginosar2021locally}, prompted by their discovery of non-periodic encodings by grid cells of three-dimensional space (see discussion), present a parsimonious model that explains both 3D and 2D representations. They model grid fields as particles that repulse each other at short distances and attract at intermediate, the dynamics then pushes the particles towards lower energy states, and the optimal state is a dense packing. Matching neural observations, running these dynamics in 2D leads to dense packing hexagonal lattices, while in 3D it often leads to jammed sub-optimal solutions without global periodic structure. 

A slightly related idea appears in \cite{huber2025memory}. In this memory model the classic roles of place and grid cells are reversed, place cells encode where a memory happens (a conjunction of a thing and a place) while grid cells encode the thing that is happening. Grid tuning curves are produced by arguing that the grid cell is encoding a variable that is uniform across space. The model then assumes that inputs that are nearby in space will be grouped into the same memory, while those beyond a critical distance will trigger a new memory. These dynamics lead to a hexagonal lattice receptive field, which can be understood via dense packing.

Despite the elegant simplicity of these approaches, simple functional questions remain non-obvious and key phenomena unexplained. Most pertinently for our current argument, no approach naturally incorporates the translational symmetry of a grid module: in Mok \& Love or Huber it is not obvious why grid cells would code for a translated version of either the conceptual fit to data or a set of memories, while in Ginosar et al. some mechanism would be required to align these densely packing lattices across neurons. Similarly unclear is why there are modules with a discrete set of lengthscales or conjunctive grid cells. Finally, why we should think of grid cells as a measure of hippocampal fit, as a discretised version of a uniform variable, or as a set of repulsing particles, when more compelling narratives exist is unclear. Nonetheless, in conjunction with other ideas, dense packing does explain the choice of hexagonal lattice in many models \citep{stemmler2015connecting,dorrell2023actionable}. 

\section{Efficient Coding Metric Loss with Large Lengthscale Produces Place Cells}

\begin{figure}[!h]
    \centering
    \includegraphics[width=0.5\linewidth]{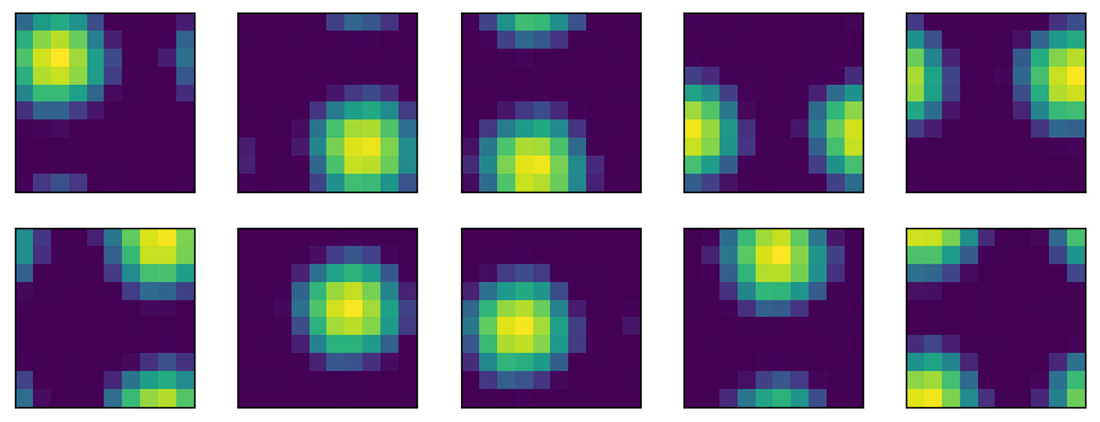}
    \caption{We optimise a metric encoding loss,~\cref{eq:pettersen} with large $\sigma$ and find the optimal representation is place cells, matching the correspondance with the similarity matching objective,~\cref{sec:grids-lit_sengupta}. We use a periodic environment for convenience, hence the multiple patches observed correspond to parts of the same field.}
    \label{fig:Pettersen_Largesig_Place}
\end{figure}
\end{document}